\documentclass[sigconf]{acmart}

\usepackage{multirow} 
\usepackage{arydshln} 
\usepackage{booktabs} 
\usepackage{amsmath}
\usepackage{makecell}
\usepackage{xcolor}
\usepackage[table]{xcolor}
\usepackage{enumitem}
\usepackage[ruled,vlined]{algorithm2e}
\usepackage[dvipsnames]{xcolor}

\renewcommand\footnotetextcopyrightpermission[1]{}
\newcommand{\ModelName}{$\mathrm{PruneRAG}$}

\AtBeginDocument{%
  }

\setcopyright{acmlicensed}
\copyrightyear{2018}
\acmYear{2018}
\acmDOI{XXXXXXX.XXXXXXX}
\acmConference[Conference acronym 'XX]{Make sure to enter the correct
  conference title from your rights confirmation email}{June 03--05,
  2018}{Woodstock, NY}
\acmISBN{978-1-4503-XXXX-X/2018/06}

\begin{document}


\title{PruneRAG: Confidence-Guided Query Decomposition Trees for Efficient Retrieval-Augmented Generation}



\author{Shuguang Jiao}
\affiliation{
    \institution{Harbin Institute of Technology, Shenzhen}
    \city{Shenzhen}
    \country{China}
}
\email{24S151028@stu.hit.edu.cn}

\author{Xinyu Xiao}
\affiliation{
    \institution{Harbin Institute of Technology, Shenzhen}
    \city{Shenzhen}
    \country{China}
}
\email{23b951021@stu.hit.edu.cn}

\author{Yunfan Wei}
\affiliation{
    \institution{South China University of Technology}
    \city{Guangzhou}
    \country{China}
}
\email{msyunfan@mail.scut.edu.cn}

\author{Shuhan Qi}
\authornote{Corresponding authors.}
\affiliation{
    \institution{Harbin Institute of Technology, Shenzhen and Leanplans}
    \city{Shenzhen}
    \country{China}
}
\email{shuhanqi@cs.hitsz.edu.cn}

\author{Chengkai Huang}
\authornotemark[1] 
\affiliation{
    \institution{Macquarie University and UNSW}
    \city{Sydney}
    \country{Australia}
}
\email{chengkai.huang1@unsw.edu.au}

\author{Quan Z. Michael Sheng}
\affiliation{
    \institution{Macquarie University}
    \city{Sydney}
    \country{Australia}
}
\email{michael.sheng@mq.edu.au}

\author{Lina Yao}
\affiliation{
    \institution{UNSW and CSIRO’s Data61}
    \city{Sydney}
    \country{Australia}
}
\email{lina.yao@unsw.edu.au}

\renewcommand{\shortauthors}{Shuguang et al.}

\begin{abstract}

Retrieval-augmented generation (RAG) has become a powerful framework for enhancing large language models in knowledge-intensive and reasoning tasks. However, as reasoning chains deepen or search trees expand, RAG systems often face two persistent failures: evidence forgetting, where retrieved knowledge is not effectively used, and inefficiency, caused by uncontrolled query expansions and redundant retrieval. These issues reveal a critical gap between retrieval and evidence utilization in current RAG architectures. We propose \ModelName{}, a confidence-guided query decomposition framework that builds a structured query decomposition tree to perform stable and efficient reasoning. PruneRAG introduces three key mechanisms: adaptive node expansion that regulates tree width and depth, confidence-guided decisions that accept reliable answers and prune uncertain branches, and fine-grained retrieval that extracts entity-level anchors to improve retrieval precision. Together, these components preserve salient evidence throughout multi-hop reasoning while significantly reducing retrieval overhead. To better analyze evidence misuse, we define the Evidence Forgetting Rate as a metric to quantify cases where golden evidence is retrieved but not correctly used. 
Extensive experiments across various multi-hop QA benchmarks show that \ModelName{} achieves superior accuracy and efficiency over state-of-the-art baselines.
The code is publicly available.\footnote{\url{https://github.com/Fdioa/PruneRAG}}
\end{abstract}




\received{20 February 2007}
\received[revised]{12 March 2009}
\received[accepted]{5 June 2009}

\maketitle





\section{Introduction}

\begin{figure}[t]
    \centering
\includegraphics[width=0.8\linewidth]{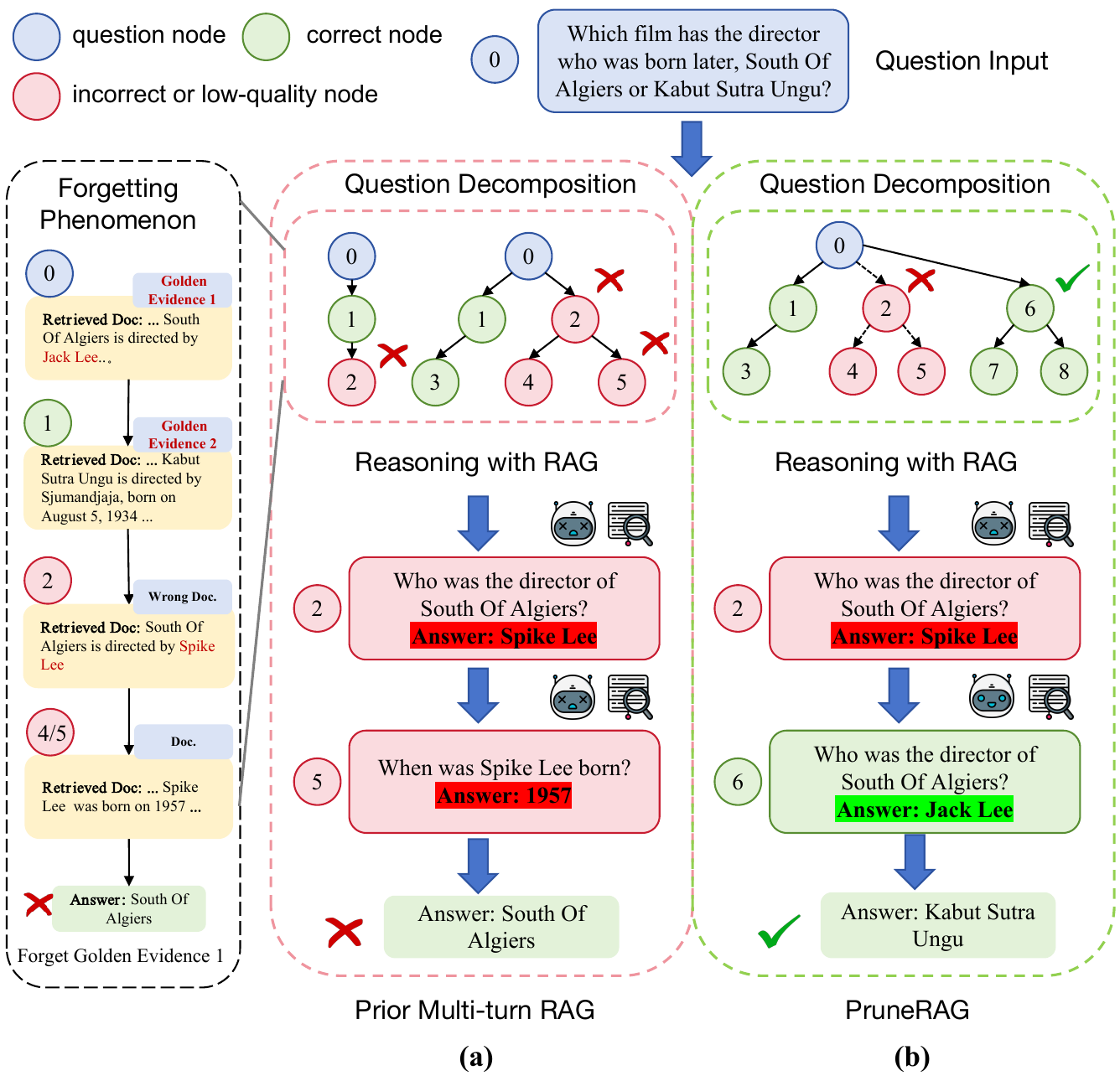}
    \vspace{-1.5em}
    \caption{The left Figure (a) illustrates Multi-turn RAG's sequential decomposition, showing how a low-quality or incorrect answer from an intermediate sub-question propagates, leading to an incorrect final result.
    The right Figure (b) demonstrates our PruneRAG method. By rejecting the low-quality answer and reflecting to generate a correct, high-quality answer, PruneRAG prevents downward error propagation and successfully arrives at the correct final answer.}
    \label{fig1}
\end{figure}

Retrieval-augmented generation (RAG) provides an effective and widely adopted paradigm for enhancing Large Language Models (LLMs) on knowledge-intensive and reasoning tasks \cite{lewis2020retrieval, guu2020retrieval, DBLP:conf/emnlp/shuster2021retrieval,huang2025embedding,huang2024foundation}. 
In open-domain QA and multi-hop fact verification, LLMs need to integrate evidence across multiple documents, often via iterative retrieval and reasoning. 
A common strategy is to decompose complex questions into sub-questions and perform step-by-step retrieval and generation \cite{DBLP:conf/emnlp/probtree_cao, li2025search}.

However, as reasoning chains grow longer or search trees deepen, two common failures emerge: 
\textit{evidence forgetting}, where the model retrieves key evidence but fails to leverage it in later steps, and 
\textit{inefficiency}, where uncontrolled expansions and redundant retrievals lead to high latency and cost. 
Figure~\ref{fig1}(a) illustrates the forgetting phenomenon on HotpotQA, where golden evidence is retrieved early but later overshadowed by accumulated context. 
These failures reveal a persistent and critical gap between retrieval and effective evidence utilization in current RAG systems.

\textbf{Chain-based RAG} extends sequential reasoning to retrieval-augmented settings \cite{trivedi2022ircot, yao2023react, chen2025towards}, decomposing a complex question into sub-questions that are solved step by step while dynamically invoking external knowledge \cite{DBLP:conf/emnlp/fu2021decomposing}. 
As question complexity increases, reasoning chains lengthen and crucial early evidence is gradually diluted or forgotten over successive steps \cite{wang2024forgetting}, destabilizing the final answers. 
Although on-demand retrieval reduces part of the redundancy, the lack of strong filtering and selection allows irrelevant content to be introduced, further exacerbating the forgetting of key information \cite{liu2023lost, singh2025chunkrag}.

\textbf{Tree-based RAG} seeks robustness by exploring multiple reasoning paths and selecting a promising trajectory \cite{roy-etal-2024-contregen, DBLP:conf/acl/peng2025grat, DBLP:conf/coling/shen2025reasoningtree, DBLP:conf/acl/sun2025era, jiang-etal-2025-rag}. 
For example, ConTReGen hierarchically expands semantic facets of a query \cite{roy-etal-2024-contregen}, and RAG-Star employs Monte Carlo tree search to navigate paths \cite{jiang-etal-2025-rag}. 
Yet as tree depth increases, these methods struggle to maintain effective transmission of salient evidence: key facts are mixed with redundant context, and the absence of disciplined node-expansion and pruning yields many low-value nodes, driving up inference cost and obscuring crucial information.

Across both families, a common weakness is the lack of a \textit{confidence-guided control mechanism} that decides when to stop answering, when to split a query, and when to refine retrieval. 
Without reliable control, reasoning tends to over-expand or drift, resulting in a systemic pattern of “retrieved but unused” evidence. 

We propose \ModelName{}, a tree-structured RAG framework for stable and efficient parallel reasoning as shown in Figure~\ref{fig1}(b). 
\ModelName{} iteratively constructs a query decomposition tree that breaks a complex question into independently solvable sub-problems, then aggregates child information during a backtracking phase to preserve and transmit key evidence throughout the reasoning process. 
To further improve efficiency and robustness, \ModelName{} integrates three mechanisms: 
(i) \emph{adaptive node expansion}, which regulates width and depth by deciding expansion based on the current query and retrieved context; 
(ii) \emph{confidence-guided decisions}, which estimate answer confidence from token-level probabilities to accept high-confidence answers early and suppress low-confidence branches that cause drift; and 
(iii) \emph{fine-grained retrieval}, which extracts key entities when a query cannot be further decomposed, enabling precise retrieval that counteracts noise and context mismatch.

To diagnose the core failure of evidence misuse, we introduce the \textbf{Evidence Forgetting Rate (EFR)}, which measures the proportion of cases where golden evidence has been retrieved but the answer is still incorrect. 
EFR complements standard end-task accuracy by directly targeting evidence utilization; we provide the formal definition in Section~\ref{evaluation}.

The main contributions of this paper are as follows:

\begin{itemize}[leftmargin=1em, itemindent=0em, itemsep=0.25em, topsep=0.25em]
    \item We propose \ModelName{}, a confidence-guided query decomposition tree for RAG that unifies answering, decomposition, and fine-grained retrieval, enabling structured and parallelizable reasoning while mitigating key-evidence forgetting.
    \item We design an adaptive expansion and pruning mechanism guided by answer probabilities, improving efficiency and reducing both redundancy and path drift.
    \item We introduce the evidence forgetting rate (EFR) as a diagnostic metric of evidence usability, and show that \ModelName{} consistently reduces EFR across datasets, with an average reduction of 20.8\% compared to mainstream multi-retrieval baselines, demonstrating its effectiveness in mitigating evidence forgetting.
    \item We conduct evaluations on three multi-hop QA datasets. Experimental results show that \ModelName{} improves F1 score by 5.45\% on average over the strongest baseline, while achieving a 4.9$\times$ speedup compared to mainstream multi-retrieval baselines.
\end{itemize}

\section{Related Work}

\textbf{Retrieval-Augmented Reasoning.}
Retrieval-augmented generation (RAG) has become a key paradigm for improving the factuality of large language models (LLMs). Representative methods such as REALM \cite{guu2020retrieval}, RAG \cite{lewis2020retrieval}, and DPR \cite{karpukhin2020dense} incorporate document retrieval during inference to compensate for limited parametric knowledge, showing strong performance across QA, fact verification, and knowledge-intensive tasks. Meanwhile, Chain-of-Thought (CoT) \cite{wei2022chain} prompting enhances reasoning interpretability and stability by encouraging intermediate steps. Recent efforts combine CoT with RAG \cite{DBLP:conf/naacl/jeong2024adaptive,DBLP:conf/acl/su2024dragin}: ReAct \cite{yao2023react} alternates between reasoning and retrieval in a closed loop, Self-RAG \cite{DBLP:conf/iclr/AsaiWWSH24} generates intermediate queries to retrieve supporting evidence, and Search-o1 \cite{li2025search} selects and integrates relevant content at a fine-grained level. Although chain-based methods simplify the reasoning process, their inherently linear structure makes it difficult to support parallel processing of multiple reasoning paths. Moreover, since information is passed in a linearly accumulated manner between steps, critical information retrieved in earlier stages tends to be naturally diluted by later context, leading to information forgetting, reasoning path drift, and redundant retrieval—ultimately limiting the depth and efficiency of reasoning in complex tasks.

\textbf{Tree-based Retrieval-Augmented Generation.} In existing tree-based Retrieval-Augmented Generation methods \cite{yao2023tree,xu2024search,DBLP:conf/acl/sun2025era,feng2025airrag,DBLP:conf/acl/peng2025grat,DBLP:conf/emnlp/probtree_cao}, the tree structure mainly serves as a tool for information expansion and path search. ConTReGen’s \cite{roy-etal-2024-contregen} tree structure achieves retrieval expansion by hierarchically exploring multiple semantic facets of queries, but it does not implement multi-step reasoning to solve problems and only expands retrieval knowledge; RAG-Star \cite{jiang-etal-2025-rag} leverages Monte Carlo Tree Search to explore reasoning paths, but its external knowledge is only used to verify intermediate reasoning results and does not participate in the generation of reasoning nodes. Essentially, it is an optimal search of linear reasoning paths and cannot support independent parallel reasoning. Moreover, due to the lack of efficient structural control mechanisms, existing methods tend to generate a large number of redundant nodes and retrieved documents when handling complex queries, leading to prolonged reasoning processes and significantly increased computational overhead \cite{shen2024understanding,DBLP:conf/sigir/salemi2024evaluating}, which severely limits their applicability in high-efficiency reasoning scenarios. 


However, existing methods do not explicitly address the issue of evidence forgetting, where retrieved golden information fails to be effectively utilized in subsequent reasoning steps.

\section{PruneRAG}

In this section, we introduce \ModelName{}, which integrates three key mechanisms: adaptive query decomposition (Section~\ref{sec:ada}) for structured problem decomposition, confidence-guided pruning (Section~\ref{sec:pruning}) for filtering low-confidence branches and avoiding redundancy, and fine-grained retrieval (Section~\ref{sec:fine-grained}) for precise evidence acquisition when decomposition fails. Together, these mechanisms enhance reasoning accuracy, stability, and efficiency.

\begin{figure*}[t]
    \centering
    \includegraphics[width=\textwidth]{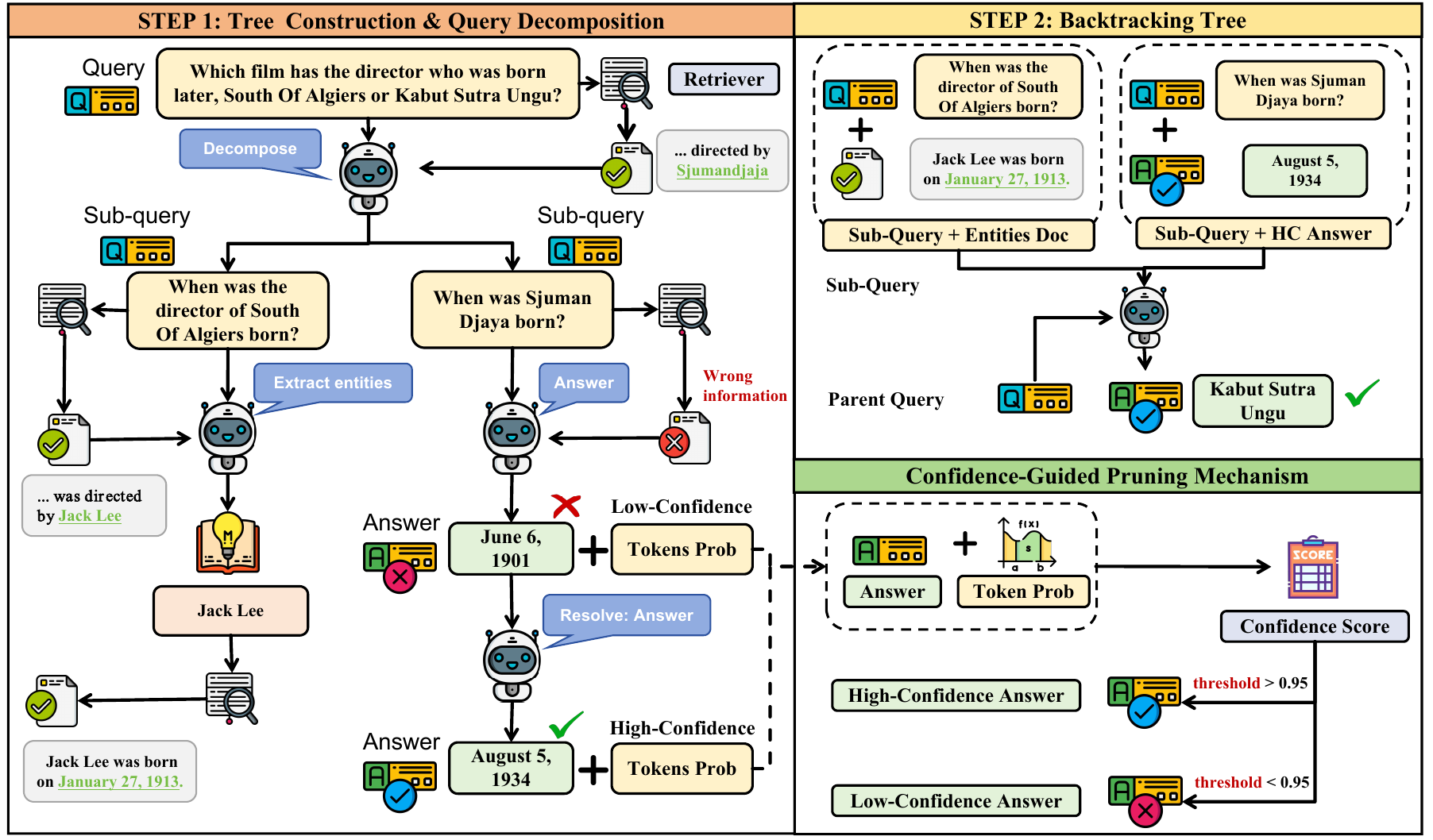} 
    \caption{Overall framework of our proposed PruneRAG. The model constructs a query decomposition tree via confidence-guided expansion (Section~\ref{sec:ada}) and pruning (Section~\ref{sec:pruning}), performs fine-grained retrieval when decomposition is infeasible (Section~\ref{sec:fine-grained}), and aggregates intermediate results through a bottom-up backtracing process to generate the final answer (Section~\ref{sec:backtracing}).}
    \Description{}
    \label{fig2}
\end{figure*}

\subsection{Tree Structure Description}

We categorize nodes based on their functional roles and behavioral dynamics during recursive query processing. These node types not only determine how the decomposition tree is expanded during top-down hierarchical reasoning, but also govern how results are aggregated during bottom-up inference. We define two primary node types as follows:

\subsubsection{Query Node}

Solving a complex multi-hop question involves a sequence of reasoning steps, where each step poses a sub-question conditioned on the information obtained from earlier answers. In this paper, the sub-questions correspond to the sub-queries that guide the retrieval and reasoning.
Thus, we model the overall reasoning process as a query decomposition tree, where each node $N$ acts as an execution unit for a specific sub-task. This structure allows the system to modularize complex questions into smaller components. 
Formally, a query node is defined as:
\begin{equation}
N_q = (q, d, a),
\end{equation}

where $q$ is the sub-query associated with the current node, $d$ is the relevant retrieved document set based on $q$, and $a$ is the candidate generated answer using the provided context $d$.

Query nodes represent the fundamental abstraction in our tree and are responsible for handling both the initial input query and any intermediate sub-queries generated during decomposition. If the model produces a high-confidence answer after querying the decomposition module, the answer is stored in $a$, and this node becomes a leaf, halting further expansion. If the node is decomposed into sub-query nodes, each child is used to perform retrieval and reasoning, and the answer field is left empty until a bottom-up aggregation is performed. Similarly, if entity nodes are generated instead, the answer field is also deferred until backtracking.

\subsubsection{Entity Node}

Entity nodes handle fallback cases where a query is neither directly answerable nor further decomposable. In such scenarios, the system abstracts the query into key semantic entities and uses them for targeted evidence retrieval. This abstraction enables LLM to continue reasoning over meaningful units even when a complete answer or decomposition is not viable. Formally, each entity node is defined as:
\begin{equation}
N_e = (e, d),
\end{equation}
where $e$ is a tuple of salient entities extracted from the parent query, and $d$ is the set of documents retrieved using them.

Entity nodes are terminal by design and do not spawn further children. They enhance the flexibility and robustness of the query tree by separating entity-level evidence collection from higher-level reasoning, particularly in cases involving vague, ambiguous, or semantically entangled queries.

\subsection{Query Decomposition Module}

This module, denoted as $\texttt{f\_decompose}(q, d, p)$, constitutes a core functional process of our framework. It determines, given the current query $q$, its retrieved context $d$, and the parent query $p$, whether the query should be directly answered, decomposed into simpler sub-queries, or abstracted into entities for fine-grained retrieval.

\subsubsection{Adaptive Node Expansion Mechanism}  
\label{sec:ada}
In the proposed query decomposition module, the system employs a top-down hierarchical decision process for each input query to determine its evolution path within the decomposition tree dynamically. This mechanism consists of three progressive judgment layers: first, the system assesses whether the current query can generate a reliable answer directly based on the retrieved evidence and the parent query context. If the answer is clear and exhibits high confidence, the current node is labeled as type \texttt{"answer"}, indicating that the branch has converged and requires no further expansion.

If the query cannot be answered directly, the system further evaluates its decomposability—whether it can be rewritten as two more fundamental sub-queries. If feasible, a \texttt{"query"}-type intermediate node is constructed, and the sub-queries are processed recursively. When the query is unanswerable and undecomposable, the system executes a semantic abstraction strategy by extracting key entities from the query and constructing an \texttt{"entity"}-type node, thus terminating the branch as a structured semantic unit. 

Both \texttt{"answer"} and \texttt{"query"} nodes are considered query nodes, where \texttt{"answer"} specifically denotes that the query has been resolved. The overall function can be formally described as follows:
\begin{equation}
\begin{aligned}
&\texttt{f\_decompose}(q, d, p) = \\
&\quad
\begin{cases}
(\texttt{"answer"}, A) & \text{if } \texttt{Ans}(q) \\
(\texttt{"query"}, q_1, q_2) & \text{if } \neg \texttt{Ans}(q) \land \texttt{Spl}(q) \\
(\texttt{"entity"}, e_1, e_2) & \text{if } \neg \texttt{Ans}(q) \land \neg \texttt{Spl}(q),
\end{cases}
\end{aligned}
\end{equation}

where $A$ is the generated answer written to the node (if available), $q_1, q_2$ are auto-generated sub-queries, $e_1, e_2$ are entities extracted from $q$, and $p$ is parent node, which provides semantic supervision to preserve goal alignment. The predicate $\texttt{Ans}(q)$ determines whether $q$ is answerable with the current context, and $\texttt{Spl}(q)$ checks decomposability. Specifically, $\texttt{Ans}(q)$ refers to the LLM's judgment as to whether the context retrieved for the current sub-query q is sufficient to support an answer to that sub-query. $\texttt{Spl}(q)$ refers to the LLM's judgment, based on the inherent logic of the problem, as to whether the problem can be decomposed into two sub-problems.

\subsubsection{Confidence-Guided Pruning Mechanism}
\label{sec:pruning}
To further mitigate the hallucination problem \cite{roy-etal-2024-contregen} that may arise when generating answer-type nodes, we design a confidence-guided decision mechanism based on the model's prediction certainty. Specifically, if the adaptive node expansion mechanism generates a candidate answer $A$, the confidence-guided decision mechanism is triggered. This mechanism computes a confidence score based on the token-level log probability of the generated answer sequence and determines whether to accept the answer. In generative question answering, LLMs typically produce an answer sequence in an autoregressive manner, predicting one token at a time. Each token $a_i$ is generated conditioned on the preceding tokens, the input query $q$, and the retrieved context $d$, following the probability distribution $P(a_i \mid a_{<i}, q, d)$. When the overall token-level log probability of the answer sequence is high, it indicates that the model is relatively confident in the generation process, semantically reliable, and syntactically coherent. Thus, we define answer confidence as:
\begin{equation}
\label{eq:main}
\texttt{Confidence}(A) = \exp\left( 
    \frac{1}{|A|} \sum_{i=1}^{|A|} \log P(a_i \mid a_{<i}, q, d) 
\right),
\end{equation}
where $A = (a_1, a_2, \dots, a_{|A|})$ denotes the ordered sequential sequence of tokens constituting the answer; $P(a_i \mid a_{<i}, q, d)$ is the conditional likelihood probability of token $a_i$ given the preceding generated tokens $a_{<i}$ along with the query and retrieved context.

This confidence score serves as a principled metric to guide node expansion decisions within the decomposition process, functioning as a pruning criterion to filter low-confidence branches early and thus improve computational efficiency and answer quality. The operational workflow for each node $N_i = (q_i, d_i, a_i)$ is as follows:

\begin{enumerate}[leftmargin=*]
    \item  Compute the confidence score of the candidate answer $A$ as $\texttt{Confidence}(A)$;
    \item If $\texttt{Confidence}(A)$ is greater than the threshold $\tau_A$, the decision mechanism accepts the answer, meaning that the model can generate a high-confidence answer based on the current information. The node is then labeled as type \texttt{"answer"}, $A$ is written into the answer field of the node, and further expansion is stopped;
    \item If $\texttt{Confidence}(A)$ is less than the threshold $\tau_A$, the decision mechanism rejects the answer. The model then autonomously decides whether to continue decomposing the query into two sub-queries or to stop decomposition and extract key entities: if decomposable, a \texttt{"query"}-type node is created and $q_1, q_2$ are generated; if not, the fine-grained retrieval mechanism is triggered, an \texttt{"entity"}-type node is generated, entities are extracted, and this branch is terminated.
\end{enumerate}

The confidence-guided decision mechanism takes the generation confidence of candidate answers as the core criterion, enhancing the system's ability to assess answer quality. It plays a key role in controlling node convergence and preventing premature termination. By rejecting low-confidence answers, it effectively mitigates hallucinations, reduces invalid expansions, and improves the stability and efficiency of the query decomposition tree, thereby improving the reliability and controllability of multi-hop reasoning.

\subsubsection{Fine-grained Retrieval Mechanism}
\label{sec:fine-grained}

When a query can no longer be decomposed, it often indicates its internal logical structure has been compressed or implicitly embedded. Further decomposition may compromise semantic integrity, introduce erroneous components, and reduce reasoning efficiency. Therefore, the system treats the query as the minimal semantic unit and terminates further decomposition. It then activates the fine-grained retrieval mechanism, which extracts key entities (such as person names, locations, and events) to construct structured retrieval anchors, guiding the system to locate relevant documents or passages within the knowledge corpus. For instance, given the query "Who is Danny Welch's employer?", it extracts "Danny Welch" and "Employer of Danny Welch". These entities are output as structured semantic information, labeling the node type as \texttt{"entity"}. The entity set is used for retrieval, and the retrieved content is inserted into the node's context field $d$.

This mechanism enables precise information acquisition by treating entities as the smallest meaningful semantic unit when structural reasoning becomes ineffective. It preserves semantic completeness, significantly enhances retrieval specificity and accuracy, and stabilizes the overall reasoning process.

\subsection{Bottom-Up Tree Backtracing}
\label{sec:backtracing}
After the query decomposition tree is fully constructed, the system performs a systematic bottom-up backtracing process to progressively and coherently aggregate evidence and answers until the root node ultimately converges to the final reliable solution. We formally define the backtracing result of a node $N$ as:
\begin{equation}
\texttt{Backtrace}(N) = f(N),
\end{equation}
where the function $f(N)$ behaves differently depending on the node type.  
Specifically:  
(i) if $N=(q,d,a)$ and the node already stores a high-confidence answer $a$, then $\texttt{Backtrace}(N)=a$;  
(ii) if $N=(q,d,\varnothing)$ is a query node without a direct answer, then $\texttt{Backtrace}(N)$ is obtained by aggregating the results of its children;  
(iii) if $N=(e,d)$ is an entity node, then $\texttt{Backtrace}(N)$ summarizes the retrieved documents $d$ into supportive evidence.  

This backtracing mechanism aggregates evidence, helping to alleviate information forgetting and improve the accuracy.

\subsection{PruneRAG Inference}

Algorithm~\ref{alg:prunerag} outlines the inference process of \textsc{PruneRAG}. Given a complex query $q$, the model first initializes the root of a query decomposition tree and places it into a processing queue.  

In the top-down construction phase, nodes are expanded until the queue is exhausted. For each node, the system retrieves the top-$k$ documents and prompts the LLM to generate a candidate answer along with its confidence. If the confidence exceeds the threshold $\tau_A$, the node is marked as an \texttt{answer} leaf and its branch terminates. Otherwise, the node is either decomposed into sub-queries, which are added to the queue, or converted into an \texttt{entity} node when further decomposition is not feasible. This dynamic expansion enables confidence-guided pruning, effectively suppressing unreliable or redundant branches.  
Once the tree is constructed, a bottom-up backtracing phase is performed. Leaf nodes either return high-confidence answers or summarized evidence from entity retrieval. Internal nodes recursively aggregate the results of their children, preserving semantic dependencies across sub-queries. Finally, the root node integrates all collected information to output the final answer $a^*$.  

This procedure ensures efficient inference: pruning reduces unnecessary exploration, while the backtracing process guarantees that evidence is aggregated coherently across the reasoning path.

\begin{algorithm}[t]
\caption{\textsc{PruneRAG} Inference}
\label{alg:prunerag}
\KwIn{Complex query $q$, Retriever $\mathcal{R}$, LLM $\mathcal{M}$, Document collection $\mathcal{D} = {d_1, \dots, d_N}$}
\KwOut{Final answer $a^*$}

\textbf{Step 1: Initialize Root Node} \\
Create root node $N_q = (q, d=\emptyset, a=\emptyset)$ and push into queue $\mathcal{Q}$.

\textbf{Step 2: Top-Down Tree Construction} \\
\While{$\mathcal{Q}$ not empty}{
  Pop a node $N=(q,d,a)$ from $\mathcal{Q}$\;
  Retrieve top-$k$ documents $d$ using $\mathcal{R}$ given $q$\;
  $\mathcal{M}(q,d) \rightarrow (A, \texttt{Confidence}(A) \eqref{eq:main})$\;

  \uIf{$\texttt{Confidence}(A) \geq \tau_A$}{
    Mark $N$ as \texttt{answer} node with $a=A$; continue\;
  }
  \uElseIf{query $q$ is decomposable}{
    Create child nodes $\{q_1, q_2\}$ and push into $\mathcal{Q}$\;
  }
  \Else{
    Create entity node $N_e=(e,d)$ from extracted $e$\;
  }
}

\textbf{Step 3: Bottom-Up Backtracing} \\
\While{there exist unprocessed internal nodes}{
  \uIf{leaf $N$ is \texttt{answer}-node}{return stored $a$ to its parent}
  \uElseIf{leaf $N$ is \texttt{entity}-node}{summarize retrieved docs $d$ as evidence}
  \Else{aggregate results of all child nodes and pass upward}
}
Return the aggregated answer $a^*$ at the root.
\end{algorithm}

\begin{table*}[t]
\centering
\caption{Comparison of answer accuracy, inference latency, and evidence forgetting rate across RAG methods on multi-hop QA datasets. Bold indicates the best result, and \underline{underline} indicates the second-best. '-' denotes that the Evidence Forgetting Rate (EFR) is not applicable: for Vanilla, no retrieval documents are used; for RAG and MemoRAG, the denominator case where all golden documents are retrieved is zero. Inference latency and EFR are compared only among multi-turn retrieval methods.}
\vspace{-1em}
\setlength{\tabcolsep}{1mm}
\begin{tabular}{c c c c c c c c c c c c}
\toprule
\multirow{2}{*}{\textbf{Types}} & 
\multirow{2}{*}{\textbf{Methods}} & 
\multirow{2}{*}{\makecell{\textbf{Time Cost} \\ (ms)}} & 
\multicolumn{3}{c}{\textbf{HotpotQA}} & 
\multicolumn{3}{c}{\textbf{2WikiQA}} & 
\multicolumn{3}{c}{\textbf{MusiQue}} \\
 \cmidrule(lr){4-6} \cmidrule(lr){7-9} \cmidrule(lr){10-12} 
& & & \textbf{EM} $\uparrow$ & \textbf{F1} $\uparrow$ & \textbf{EFR} $\downarrow$ & \textbf{EM} $\uparrow$ & \textbf{F1} $\uparrow$ & \textbf{EFR} $\downarrow$ & \textbf{EM} $\uparrow$ & \textbf{F1} $\uparrow$ & \textbf{EFR} $\downarrow$\\
\midrule
\rowcolor{gray!20} \multicolumn{12}{c}{\textit{Llama-3.1-8B-Instruct}} \\
\textit{No-Retrieval} & Vanilla & 56 & 24.2 & 30.1 & - & 19.4 & 24.8 & - & 5.4 & 9.1 & - \\
\midrule
\multirow{2}{*}{\textit{Single-Retrieval}} & Standard RAG & 153 & \underline{49.0} & \underline{55.1} & 27.6 & 23.2 & 29.9 & 47.5 & 7.2 & 11.5 & - \\
& MemoRAG & 8,566 & 31.2 & 38.8 & 53.1 & 23.0 & 28.7 & 68.1 & 7.8 & 12.3 & - \\
\midrule
\multirow{7}{*}{\textit{Multi-Retrieval}} & React & \underline{1,500} & 24.2 & 27.1 & 65.2 & 15.0 & 19.4 & 73.0 & 7.8 & 11.6 & 68.4 \\
& Search-o1 & 1,696 & 35.4 & 39.4 & 53.1 & 22.0 & 25.4 & 63.3 & 9.2 & 13.6 & 75.6 \\
& Self-RAG & 3,200 & 30.4 & 37.7 & 57.4 & 13.0 & 23.4 & 89.9 & 8.2 & 14.3 & 83.3 \\
& ConTReGen & 2,156 & 43.4 & 49.8 & \underline{35.9} & 18.6 & 23.5 & \underline{56.4} & 10.4 & 14.6 & \underline{51.6} \\
& RAG-Star & 8,216 & 45.0 & 54.6 & 51.3 & \underline{30.0} & \underline{35.2} & 74.7 & \underline{13.0} & \underline{18.9} & 78.5 \\
& ProbTree & 3,838 & 32.2 & 39.3 & 54.4 & 27.8 & 32.6 & 62.6 & 11.0 & 17.1 & 91.6 \\
& \ModelName{} (Ours) & \textbf{474} & \textbf{52.8} & \textbf{60.6 {\color{ForestGreen}(5.5$\uparrow$)}} & \textbf{32.8} & \textbf{33.0} & \textbf{40.2 {\color{ForestGreen}(5.0$\uparrow$)}} & \textbf{51.4} & \textbf{15.4} & \textbf{22.9 {\color{ForestGreen}(4.0$\uparrow$)}} & \textbf{52.0} \\
\midrule
\rowcolor{gray!20} \multicolumn{12}{c}{\textit{Qwen3-8B}} \\
\textit{No-Retrieval}& Vanilla & 712 & 26.2 & 34.1 & - & \underline{31.2} & 36.1 & - & 9.0 & 15.3 & - \\
\midrule
\multirow{2}{*}{\textit{Single-Retrieval}} & Standard RAG & 824 & \underline{49.0} & 55.6 & 24.8 & 23.2 & 27.6 & 27.5 & 10.8 & 16.4 & -  \\
& MemoRAG & 10,321 & 39.4 & 50.9 & 35.1 & 22.8 & 32.1 & 65.9 & 9.0 & 16.1 & - \\
\midrule
\multirow{7}{*}{\textit{Multi-Retrieval}}  & React & 4,497 & 30.0 & 36.7 & 46.4 & 23.6 & 28.7 & 54.2 & 10.2 & 15.7 & 69.5 \\
& Search-o1 & 4,841 & 36.4 & 42.6 & 56.9 & 23.2 & 27.5 & 54.3 & \underline{12.4} & 17.4 & 62.1 \\
& Self-RAG & 3,200 & 30.4 & 37.7 & 57.4 & 13.0 & 23.4 & 89.9 & 8.2 & 14.3 & 85.7 \\
& ConTReGen & 4,507 & 47.6 & 55.4 & \underline{30.5} & 29.0 & 33.9 & \underline{27.8} & 11.8 & 17.2 & \underline{60.0}\\
& RAG-Star & 14,709 & 48.6 & \underline{57.8} & 52.4 & 31.0 & \underline{37.8} & 73.9 & 12.2 & \underline{19.8} & 84.2 \\
& ProbTree & \underline{3,300} & 22.6 & 26.4 & 50.0 & 15.8 & 19.2 & 50.6 & 5.0 & 7.6 & 84.3\\
& \ModelName{} (Ours) & \textbf{2,254} & \textbf{56.6} & \textbf{63.6 {\color{ForestGreen}(5.8$\uparrow$)}} & \textbf{23.1} & \textbf{41.2} & \textbf{44.4 {\color{ForestGreen}(6.6$\uparrow$)}} & \textbf{26.0} & \textbf{17.6} & \textbf{25.6 {\color{ForestGreen}(5.8$\uparrow$)}} & \textbf{38.4} \\
\bottomrule
\end{tabular}
\label{tab:main_results}
\end{table*}

\section{Experiments}

\subsection{Research Questions}

In this section, we conduct a comprehensive experimental study to address the following research questions (RQs):

\begin{itemize}[leftmargin=1em, itemindent=0em, itemsep=0.3em]
  \item \textbf{(RQ1)} Does evidence forgetting occur in multi-turn RAG, and can our method mitigate it to ensure reliable answer generation?
  \item \textbf{(RQ2)} Do the individual modules, especially the confidence mechanism, contribute to accuracy improvements by alleviating evidence forgetting?
  \item \textbf{(RQ3)} Can our method maintain or improve accuracy while reducing retrieval cost through fewer yet more effective retrieval steps?
  \item \textbf{(RQ4)} How sensitive is our method to hyperparameters like confidence threshold and tree depth, and what settings yield robust performance?

\end{itemize}

\subsection{Experimental Setup}

\subsubsection{Datasets}

We conducted experiments on four representative multi-hop question answering (QA) datasets: HotpotQA \cite{DBLP:conf/emnlp/Yang0ZBCSM18}, 2WikiMultihopQA (2WikiQA) \cite{DBLP:conf/coling/HoNSA20}, Musique \cite{DBLP:journals/tacl/TrivediBKS22}, Bamboogle \cite{DBLP:conf/emnlp/PressZMSSL23} and GPQA \cite{DBLP:journals/corr/gpqa_rein}. To compare with other QA tasks, we included two single-hop QA datasets, Natural Questions (NQ) \cite{DBLP:journals/tacl/Kwiatkowski2019nq} and TriviaQA\cite{DBLP:conf/acl/Joshi2017tiviaqa}. Details can be found in the Appendix \ref{app:datasets}.

\subsubsection{Evaluation Metrics}\label{evaluation}
To evaluate the performance of our method, following previous works \cite{li2025search}, we consider two dimensions: (i) \textit{effectiveness}, measuring answer quality and coverage, and (ii) \textit{efficiency}, capturing reasoning cost in time and resources.
Effectiveness is measured by Exact Match (EM) and F1, which assess answer correctness at both strict and token-overlap levels. Efficiency is evaluated through Document Recall (Recall), reflecting whether the golden document is retrieved; Retrieval Number (RN), indicating how often the retriever is invoked; Time Cost (Time), measured as the average model inference time per question under a fixed \texttt{max\_tokens} setting; and Evidence Forgetting Rate (EFR).

To be specific, to capture the phenomenon where a model retrieves all the necessary evidence but still fails to generate the correct answer, we define the \emph{Evidence Forgetting Rate (EFR)} as:
\begin{equation}
\text{EFR} = \frac{1}{N} \sum_{i=1}^{N} \mathbf{1}\{\, G_i \subseteq d_i \;\wedge\; a_i \neq a_i^* \,\}.
\end{equation}

where $N$ denotes the total number of evaluation samples, $G_i$ is the golden evidence set containing all documents required to answer the $i$-th question, $d_i$ represents the set of documents retrieved by the model throughout all retrieval rounds for that question, $a_i$ is the model’s predicted answer, and $a_i^*$ is the corresponding ground-truth answer. The indicator function $\mathbf{1}\{\cdot\}$ takes value 1 if the retrieved evidence fully covers the golden set ($G_i \subseteq d_i$) while the generated answer is incorrect ($a_i \neq a_i^*$), and 0 otherwise. Thus, EFR quantifies the proportion of cases in which the model fails to utilize complete evidence effectively, reflecting the severity of evidence forgetting in retrieval-augmented generation.

\subsubsection{Baselines}
We compare \ModelName{} with four chain-based RAG methods: React \cite{yao2023react}, Search-o1 \cite{li2025search}, Self-RAG \cite{DBLP:conf/iclr/AsaiWWSH24} and MemoRAG \cite{DBLP:conf/www/memorag_qian}, as well as three tree-structured RAG methods: ConTRGen \cite{roy-etal-2024-contregen}, RAG-Star \cite{jiang-etal-2025-rag} and ProbTree\cite{DBLP:conf/emnlp/probtree_cao}. React integrates language model reasoning with retrieval actions in an interleaved reasoning-action process; Search-o1 incorporates a document reasoning module to summarize and integrate document knowledge into the reasoning process. Self-RAG employs reflection tokens to let the model assess retrieved content and perform adaptive, on-demand retrieval. ConTRGen explores multiple semantic facets of the query hierarchically via a tree; RAG-Star utilizes the Monte Carlo Tree Search algorithm to explore the optimal reasoning path.

\subsubsection{Implementation Details}

Following previous work \cite{jiang-etal-2025-rag},
we conduct experiments with two models: Qwen-3-8B \cite{yang2025qwen3technicalreport} and Llama-3.1-8B-Instruct \cite{grattafiori2024llama3herdmodels}. For Self-RAG, which is fine-tuned on Llama-2-7B-HF \cite{DBLP:journals/corr/llama2}, we use its official release. Following \citeauthor{karpukhin2020dense} \shortcite{karpukhin2020dense}, we use the full Wikipedia 2018 dump as the retrieval corpus. Our main experiments follow the retrieval setup of \citeauthor{jiang-etal-2025-rag} \shortcite{jiang-etal-2025-rag}, employing FAISS \cite{johnson2019billion} for indexing, BGE-large-en-v1.5 \cite{DBLP:journals/corr/xiao2023bge} as retriever, and retrieving top-5 documents for answer generation. For \ModelName, we set the tree maximum branching factor to 2, the maximum tree depth to 3, and the confidence threshold to 0.95. To ensure reproducibility, decoding is performed via greedy search with temperature 0. Experiments are conducted on a server with four NVIDIA L40 PCIe 48GB GPUs, and all inference is conducted using the vLLM framework\cite{DBLP:journals/corr/ultraled,DBLP:conf/sosp/vllm_kwon}. Further details can be found in the Appendix \ref{app:details}.

\subsection{Main Result (RQ1)}

To investigate whether evidence forgetting is a prevalent issue in RAG, we compare representative methods under three retrieval paradigms: no retrieval, single-turn retrieval, and multi-turn retrieval. The results are presented in Table~\ref{tab:main_results}.  

In the no-retrieval paradigm, the model answers questions purely from parametric knowledge, resulting in relatively low accuracy. The Evidence Forgetting Rate (EFR) is not applicable in this paradigm, since no external documents are retrieved. Under the single-turn retrieval paradigm, external knowledge is incorporated through a single retrieval and generation step. While this paradigm improves accuracy, it does not involve iterative reasoning, and thus, the phenomenon of evidence forgetting is relatively mild.  

In the multi-turn retrieval paradigm, iterative methods consistently exhibit high EFR across datasets, confirming that evidence forgetting is a common weakness of this paradigm. For instance, ReAct, Search-o1, Self-RAG, RAG-Star, and ProbTree all exceed 46\% EFR on HotpotQA, rise above 50\% on 2WikiQA, and surpass 68\% on Musique, despite the fact that golden documents have been fully retrieved. Importantly, these trends are consistent across different backbone models, indicating that evidence forgetting is a general weakness of multi-turn retrieval, independent of the base model. An exception is ConTReGen, which achieves a lighter degree of forgetting (35.9\% and 30.5\% on HotpotQA), benefiting from its tree-structured design that compresses reasoning depth. However, such mitigation remains limited, as the improvement in EFR does not translate into optimal accuracy.
In contrast, \ModelName{} consistently achieves the best or second-best EFR across all datasets under both backbone models, with values such as 23.1\% on HotpotQA, 26.0\% on 2WikiQA, and 38.4\% on Musique (all reported under the Qwen-3-8B backbone). Beyond reducing forgetting, \ModelName{} also delivers the highest EM and F1 scores, with improvements in accuracy arising from reduced evidence forgetting. At the same time, \ModelName{} achieves the fastest inference speed among multi-turn RAG methods, owing to its confidence-guided pruning mechanism that effectively reduces redundant and wrong expansions, and on average runs 4.9$\times$ faster than mainstream multi-retrieval baselines.

Overall, the results validate RQ1, showing that evidence forgetting widely exists in multi-turn retrieval methods, and that \ModelName{} provides an effective solution by significantly reducing forgetting while achieving state-of-the-art accuracy and efficiency.

\subsection{Ablation Study (RQ2)}

\begin{table}[ht]
\centering
\caption{Ablation results of PruneRAG under the Qwen3-8B backbone. 
\textbf{Bold} denotes the best results. RN indicates average retrieval number (lower is better).}
\vspace{-1em}
\setlength{\tabcolsep}{0.9mm}
\begin{tabular}{l c c c c c c c c c}
\toprule 
\multirow{2}{*}{\textbf{Method}}
& \multicolumn{3}{c}{\textbf{HotpotQA}} & 
\multicolumn{3}{c}{\textbf{2WikiQA}} & 
\multicolumn{3}{c}{\textbf{Musique}} \\
\cmidrule(lr){2-4} \cmidrule(lr){5-7} \cmidrule(lr){8-10}
 & \textbf{EM}  & \textbf{RN}   & \textbf{EFR} & \textbf{EM}   & \textbf{RN}  & \textbf{EFR} & \textbf{EM}   & \textbf{RN}  & \textbf{EFR} \\
\midrule
\ModelName{}   & \textbf{56.6} & 2.0 & \textbf{23.1} & \textbf{41.2} & 3.4 & \textbf{26.0} & \textbf{17.6} & 3.3 & \textbf{38.4}\\
\midrule 
w/o Con. & 53.4 & 1.9 & 25.1 & 39.6 & 3.3 & 26.2 & 16.2 & 3.2 & 51.2\\
w/o Ans. & 37.8 & 3.3 & 48.2 & 33.2 & 4.4 & 55.3 & 15.0 & 4.8 & 65.0\\
w/o Ent. & 53.4 & \textbf{1.7} & 24.4 & 40.0 & \textbf{2.8} & 27.8 & 15.4 & \textbf{3.1} & 53.3\\
w/o Ada. & 42.0 & 5.3 & 43.2 & 32.0 & 4.6 & 50.0 & 15.2 & 5.8 & 67.7\\
\bottomrule
\end{tabular}
\label{tab:ablation}
\end{table}

To investigate RQ2, we conduct an ablation study to assess the contribution of each core component in \ModelName{}. Four variants are designed: \textit{w/o Confidence}, which removes the confidence-guided pruning mechanism; \textit{w/o Answer}, which removes the answer branch in adaptive node expansion; \textit{w/o Entity}, which removes the fine-grained retrieval mechanism; and \textit{w/o Adaptive}, which eliminates the adaptive expansion strategy, decomposing queries unconditionally.
As shown in Table~\ref{tab:ablation}, removing any of the four modules leads to clear performance degradation, indicating that each design is indispensable for the overall effectiveness of \ModelName{}.  
(a) \textbf{w/o Confidence.} Eliminating the confidence-guided pruning mechanism leads to a increase in EFR, reflecting the model’s tendency to accept unreliable outputs and disrupt critical evidence transmission. Importantly, both \textit{w/o Answer} and \textit{w/o Adaptive} also remove this pruning mechanism, and their EFR values exhibit a similar escalation, consistent with \textit{w/o Confidence}. These results collectively underscore that confidence-based pruning is indispensable for mitigating evidence forgetting and for sustaining reliable answer accuracy. 
(b) \textbf{w/o Adaptive.} Eliminating adaptive control produces the most pronounced declines in most datasets, as uncontrolled decomposition generates redundant sub-queries and unstable reasoning structures. This underscores that adaptive expansion is essential for confidence-guided pruning.  
(c) \textbf{w/o Answer.} Removing the answer branch prevents the model from using intermediate conclusions, which destabilizes reasoning and increases retrieval calls, thus raising inference cost. The presence of this module reduces retrieval frequency and supports more efficient reasoning.  
(d) \textbf{w/o Entity.} Without entity-level retrieval, leaf nodes terminate prematurely, resulting in incomplete evidence utilization and increased forgetting. This weakens semantic coverage and lowers accuracy.  

In summary, these results answer RQ2: all four modules improve accuracy, with the confidence-guided pruning mechanism mitigating evidence forgetting, while the adaptive and answer branches enhance efficiency by reducing retrieval frequency.

\subsection{Retrieval Efficiency Analysis (RQ3)}

\begin{table}[t]
\centering
\caption{Comparison of different RAG methods in terms of average Retrieval Numbers (RN), document recall rates, and Retrieval Efficiency (RE).}
\vspace{-1em}
\setlength{\tabcolsep}{1mm}
\begin{tabular}{cccccccc}
\toprule
\multirow{2}{*}{\textbf{Method}} & 
\multicolumn{2}{c}{\textbf{HotpotQA}} & 
\multicolumn{2}{c}{\textbf{2Wiki}} & 
\multicolumn{2}{c}{\textbf{Musique}} &
\multirow{2}{*}{\textbf{RE}} \\
\cmidrule{2-3} \cmidrule{4-5} \cmidrule{6-7} 
& \textbf{RN}& \textbf{Recall} & \textbf{RN} & \textbf{Recall} & \textbf{RN} & \textbf{Recall} \\
\midrule
React      & 2.61 & 47.5 & 3.25 & 47.0 & 2.83 & 20.0 & 13.2 \\
Search-o1  & 3.09 & 54.3 & 4.35 & 53.8 & 3.68 & 25.8 & 12.3 \\
Self-RAG   & 1.89 & 54.3 & 1.96 & 32.4 & 1.92 & 9.7 & \textbf{16.7} \\
ConTReGen  & 3.76 & 58.0 & 2.81 & 36.3 & 3.87 & 13.3 & 10.5\\
RAG-Star   & 6.73 & 42.9 & 4.30 & 29.4 & 3.49 & 11.4 & 5.4\\
ProbTree   & 4.00 & 61.7 & 4.25 & 42.0 & 4.14 & 22.4 & 10.2\\
\ModelName{}    & 2.06 & 60.7 & 3.40 & 47.4 & 3.39 & 23.2 & \textbf{16.7}\\
\bottomrule
\end{tabular}
\Description{}
\label{tab_rn}
\end{table}
\begin{figure}[ht]
    \centering
    \includegraphics[width=\columnwidth]{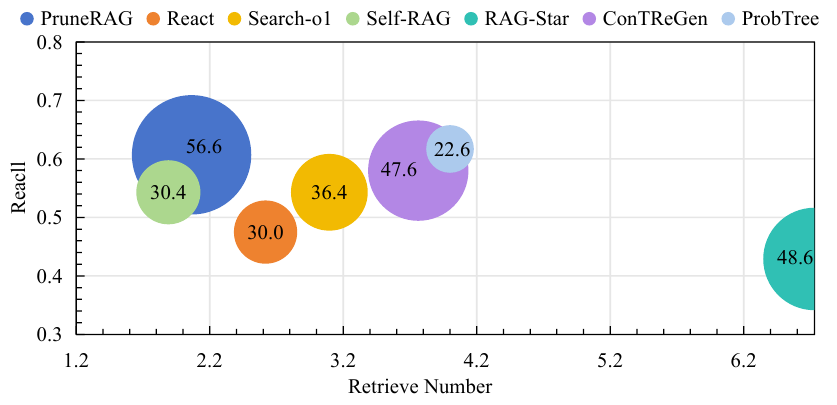} 
    \vspace{-2.5em}
    \caption{Retrieval efficiency on HotpotQA, where the x-axis denotes retrieval count, the y-axis denotes golden document recall rate, and bubble size indicates EM.}
    \Description{}
    \label{fig3}
\end{figure}

Table~\ref{tab_rn} compares document recall (Recall), the retrieval number (RN) and Retrieval Efficiency (RE) across different methods. 
RE refers to the ratio between document recall (Recall) and the retrieval number (RN). The results show that \ModelName{} consistently achieves higher recall with fewer retrieval calls. This advantage stems from dynamic query decomposition, where sub-queries are generated based on the parent query and its retrieved documents, and redundant sub-queries are suppressed once the corresponding evidence has already been obtained. Moreover, the fine-grained retrieval mechanism further enhances recall: when a query can no longer be decomposed or directly answered, key entities with reduced noise are extracted for targeted retrieval, improving the likelihood of capturing relevant evidence.

Figure~\ref{fig3} further illustrates the trade-off between retrieval count, recall rate, and accuracy on HotpotQA. While baseline methods often rely on increasing retrievals to boost recall, this strategy introduces redundant or distracting information and fails to translate into accuracy gains. In contrast, \ModelName{} achieves higher EM scores even at comparable or lower recall levels, demonstrating a superior ability to exploit retrieved golden documents effectively.

In summary, these findings answer RQ3: \ModelName{} attains higher information coverage with fewer retrievals, and its principled mechanisms enable more efficient utilization of external knowledge, thereby improving accuracy while controlling retrieval cost.

\subsection{Hyper-parameters Analysis (RQ4)}

\begin{figure}[t]
    \centering
    \includegraphics[width=1\columnwidth]{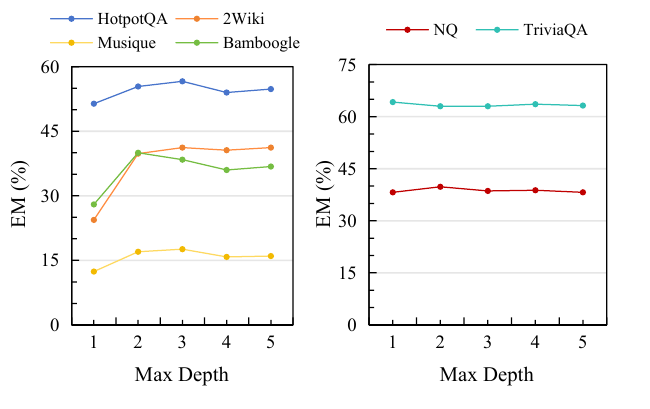} 
    \vspace{-3em}
    \caption{Impact of the maximum query tree depth on performance across datasets with varying reasoning complexity.}
    \Description{}
    \label{fig4}
\end{figure}


\begin{figure}[t]
    \centering
    \includegraphics[width=1\columnwidth]{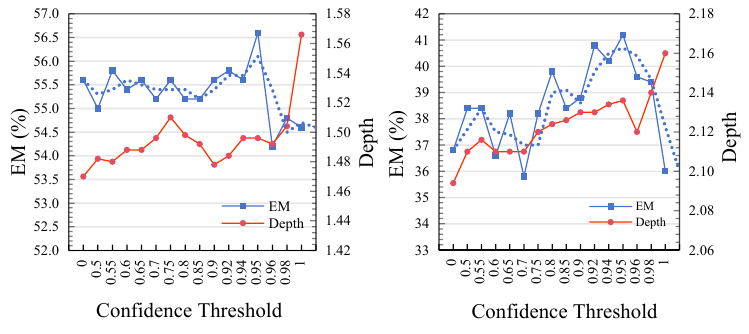} 
    \vspace{-2.5em}
    \caption{Impact of confidence threshold $\tau_A$ on answer accuracy (Exact Match) and average query tree depth across HotpotQA (left) and 2WikiQA (right). The results demonstrate that the threshold range $\tau_A \in [0.9, 0.95]$ consistently achieves strong performance.}
    \label{main_fig6}
    \Description{}
\end{figure}

To address RQ4, we analyze the sensitivity of \ModelName{} to two key hyperparameters: the maximum query tree depth and the confidence threshold $\tau_A$.  

\textbf{Maximum Tree Depth.}  
As illustrated in Figure~\ref{fig4}, \ModelName{} demonstrates a strong ability to adapt reasoning depth to task complexity. On multi-hop QA datasets such as HotpotQA, performance steadily improves as the maximum depth increases, reaching an optimal point around depth 2--3 before saturating. This pattern aligns with the typical reasoning steps required for most multi-hop QA instances, indicating that the model effectively decomposes complex questions and integrates cross-paragraph information. In contrast, for predominantly single-hop datasets such as NQ and TriviaQA, the best performance is observed when the depth is set to 1, and deeper expansions introduce noise that slightly impairs accuracy. These results highlight that \ModelName{} can perceive task complexity and dynamically adjust reasoning depth accordingly.  

\textbf{Confidence Threshold $\tau_A$.}  
Figure~\ref{main_fig6} shows the impact of $\tau_A$ on answer accuracy and average tree depth. When $\tau_A < 0.5$, the model tends to accept answers prematurely, producing shallow reasoning structures and unstable accuracy. In the mid-to-high range ($\tau_A \in [0.9, 0.95]$), the confidence-guided pruning mechanism is fully activated, rejecting more low-quality answer nodes and thereby yielding deeper reasoning paths and more reliable answers. On HotpotQA and 2WikiQA, peak performance is achieved at $\tau_A = 0.95$, with EM scores of 56.6\% and 41.2\% and corresponding depths of 1.49 and 2.13, respectively. 

Overall, \ModelName{} exhibits robustness with respect to hyperparameter choices. The results demonstrate that maximum tree depth and $\tau_A$ jointly balance reasoning complexity and accuracy, with a depth of 2--3 and a threshold of $[0.9, 0.95]$ forming a stable, transferable configuration across datasets. These findings provide a clear answer to RQ4: the performance of \ModelName{} is not overly sensitive to hyperparameter selection, and principled parameter ranges can ensure consistent and reliable performance across tasks of varying complexity.  

\section{Conclusion}
In this paper, we propose PruneRAG, a confidence-guided query decomposition tree framework to address the twin challenges of evidence forgetting and inefficiency in retrieval-augmented generation. By combining adaptive expansion, confidence-guided pruning, and fine-grained entity retrieval, PruneRAG preserves key evidence while reducing redundant reasoning. Experiments on various multi-hop QA benchmarks demonstrate consistent gains in both accuracy and efficiency, with significant mitigation of evidence forgetting.



\bibliographystyle{ACM-Reference-Format}
\bibliography{sample-base}


\begin{thebibliography}{48}


\ifx \showCODEN    \undefined \def \showCODEN     #1{\unskip}     \fi
\ifx \showISBNx    \undefined \def \showISBNx     #1{\unskip}     \fi
\ifx \showISBNxiii \undefined \def \showISBNxiii  #1{\unskip}     \fi
\ifx \showISSN     \undefined \def \showISSN      #1{\unskip}     \fi
\ifx \showLCCN     \undefined \def \showLCCN      #1{\unskip}     \fi
\ifx \shownote     \undefined \def \shownote      #1{#1}          \fi
\ifx \showarticletitle \undefined \def \showarticletitle #1{#1}   \fi
\ifx \showURL      \undefined \def \showURL       {\relax}        \fi
\providecommand\bibfield[2]{#2}
\providecommand\bibinfo[2]{#2}
\providecommand\natexlab[1]{#1}
\providecommand\showeprint[2][]{arXiv:#2}

\bibitem[Asai et~al\mbox{.}(2024)]%
        {DBLP:conf/iclr/AsaiWWSH24}
\bibfield{author}{\bibinfo{person}{Akari Asai}, \bibinfo{person}{Zeqiu Wu}, \bibinfo{person}{Yizhong Wang}, \bibinfo{person}{Avirup Sil}, {and} \bibinfo{person}{Hannaneh Hajishirzi}.} \bibinfo{year}{2024}\natexlab{}.
\newblock \showarticletitle{Self-RAG: Learning to Retrieve, Generate, and Critique through Self-Reflection}. In \bibinfo{booktitle}{\emph{{ICLR}}}. \bibinfo{publisher}{OpenReview.net}.
\newblock


\bibitem[Cao et~al\mbox{.}(2023)]%
        {DBLP:conf/emnlp/probtree_cao}
\bibfield{author}{\bibinfo{person}{Shulin Cao}, \bibinfo{person}{Jiajie Zhang}, \bibinfo{person}{Jiaxin Shi}, \bibinfo{person}{Xin Lv}, \bibinfo{person}{Zijun Yao}, \bibinfo{person}{Qi Tian}, \bibinfo{person}{Lei Hou}, {and} \bibinfo{person}{Juanzi Li}.} \bibinfo{year}{2023}\natexlab{}.
\newblock \showarticletitle{Probabilistic Tree-of-thought Reasoning for Answering Knowledge-intensive Complex Questions}. In \bibinfo{booktitle}{\emph{{EMNLP} (Findings)}}. \bibinfo{publisher}{Association for Computational Linguistics}, \bibinfo{pages}{12541--12560}.
\newblock


\bibitem[Chen et~al\mbox{.}(2025)]%
        {chen2025towards}
\bibfield{author}{\bibinfo{person}{Qiguang Chen}, \bibinfo{person}{Libo Qin}, \bibinfo{person}{Jinhao Liu}, \bibinfo{person}{Dengyun Peng}, \bibinfo{person}{Jiannan Guan}, \bibinfo{person}{Peng Wang}, \bibinfo{person}{Mengkang Hu}, \bibinfo{person}{Yuhang Zhou}, \bibinfo{person}{Te Gao}, {and} \bibinfo{person}{Wanxiang Che}.} \bibinfo{year}{2025}\natexlab{}.
\newblock \showarticletitle{Towards reasoning era: A survey of long chain-of-thought for reasoning large language models}.
\newblock \bibinfo{journal}{\emph{arXiv preprint arXiv:2503.09567}} (\bibinfo{year}{2025}).
\newblock


\bibitem[Feng et~al\mbox{.}(2025)]%
        {feng2025airrag}
\bibfield{author}{\bibinfo{person}{Wenfeng Feng}, \bibinfo{person}{Chuzhan Hao}, \bibinfo{person}{Yuewei Zhang}, \bibinfo{person}{Jingyi Song}, {and} \bibinfo{person}{Hao Wang}.} \bibinfo{year}{2025}\natexlab{}.
\newblock \bibinfo{title}{AirRAG: Activating Intrinsic Reasoning for Retrieval Augmented Generation using Tree-based Search}.
\newblock
\showeprint[arxiv]{2501.10053}~[cs.AI]


\bibitem[Fu et~al\mbox{.}(2021)]%
        {DBLP:conf/emnlp/fu2021decomposing}
\bibfield{author}{\bibinfo{person}{Ruiliu Fu}, \bibinfo{person}{Han Wang}, \bibinfo{person}{Xuejun Zhang}, \bibinfo{person}{Jun Zhou}, {and} \bibinfo{person}{Yonghong Yan}.} \bibinfo{year}{2021}\natexlab{}.
\newblock \showarticletitle{Decomposing Complex Questions Makes Multi-Hop {QA} Easier and More Interpretable}. In \bibinfo{booktitle}{\emph{Findings of the Association for Computational Linguistics: {EMNLP} 2021, Virtual Event / Punta Cana, Dominican Republic, 16-20 November, 2021}}. \bibinfo{publisher}{Association for Computational Linguistics}, \bibinfo{pages}{169--180}.
\newblock


\bibitem[Grattafiori et~al\mbox{.}(2024)]%
        {grattafiori2024llama3herdmodels}
\bibfield{author}{\bibinfo{person}{Aaron Grattafiori}, \bibinfo{person}{Abhimanyu Dubey}, \bibinfo{person}{Abhinav Jauhri}, \bibinfo{person}{Abhinav Pandey}, \bibinfo{person}{Abhishek Kadian}, \bibinfo{person}{Ahmad Al-Dahle}, \bibinfo{person}{Aiesha Letman}, \bibinfo{person}{Akhil Mathur}, \bibinfo{person}{Alan Schelten}, \bibinfo{person}{Alex Vaughan}, \bibinfo{person}{Amy Yang}, \bibinfo{person}{Angela Fan}, \bibinfo{person}{Anirudh Goyal}, \bibinfo{person}{Anthony Hartshorn}, \bibinfo{person}{Aobo Yang}, \bibinfo{person}{Archi Mitra}, \bibinfo{person}{Archie Sravankumar}, \bibinfo{person}{Artem Korenev}, {and} \bibinfo{person}{et al.}} \bibinfo{year}{2024}\natexlab{}.
\newblock \bibinfo{title}{The Llama 3 Herd of Models}.
\newblock
\showeprint[arxiv]{2407.21783}~[cs.AI]


\bibitem[Guu et~al\mbox{.}(2020)]%
        {guu2020retrieval}
\bibfield{author}{\bibinfo{person}{Kelvin Guu}, \bibinfo{person}{Kenton Lee}, \bibinfo{person}{Zora Tung}, \bibinfo{person}{Panupong Pasupat}, {and} \bibinfo{person}{Mingwei Chang}.} \bibinfo{year}{2020}\natexlab{}.
\newblock \showarticletitle{Retrieval augmented language model pre-training}. In \bibinfo{booktitle}{\emph{International conference on machine learning}}. PMLR, \bibinfo{pages}{3929--3938}.
\newblock


\bibitem[Ho et~al\mbox{.}(2020)]%
        {DBLP:conf/coling/HoNSA20}
\bibfield{author}{\bibinfo{person}{Xanh Ho}, \bibinfo{person}{Anh{-}Khoa~Duong Nguyen}, \bibinfo{person}{Saku Sugawara}, {and} \bibinfo{person}{Akiko Aizawa}.} \bibinfo{year}{2020}\natexlab{}.
\newblock \showarticletitle{Constructing {A} Multi-hop {QA} Dataset for Comprehensive Evaluation of Reasoning Steps}. In \bibinfo{booktitle}{\emph{Proceedings of the 28th International Conference on Computational Linguistics, {COLING} 2020, Barcelona, Spain (Online), December 8-13, 2020}}. \bibinfo{publisher}{International Committee on Computational Linguistics}, \bibinfo{pages}{6609--6625}.
\newblock


\bibitem[Huang et~al\mbox{.}(2020)]%
        {huang2020explore}
\bibfield{author}{\bibinfo{person}{Chengkai Huang}, \bibinfo{person}{Xuan Luo}, \bibinfo{person}{Jiajia Zhang}, \bibinfo{person}{Qing Liao}, \bibinfo{person}{Xuan Wang}, \bibinfo{person}{Zoe~L Jiang}, {and} \bibinfo{person}{Shuhan Qi}.} \bibinfo{year}{2020}\natexlab{}.
\newblock \showarticletitle{Explore instance similarity: An instance correlation based hashing method for multi-label cross-model retrieval}.
\newblock \bibinfo{journal}{\emph{Information Processing \& Management}} \bibinfo{volume}{57}, \bibinfo{number}{2} (\bibinfo{year}{2020}), \bibinfo{pages}{102165}.
\newblock


\bibitem[Huang et~al\mbox{.}(2025b)]%
        {huang2025embedding}
\bibfield{author}{\bibinfo{person}{Chengkai Huang}, \bibinfo{person}{Yu Xia}, \bibinfo{person}{Rui Wang}, \bibinfo{person}{Kaige Xie}, \bibinfo{person}{Tong Yu}, \bibinfo{person}{Julian McAuley}, {and} \bibinfo{person}{Lina Yao}.} \bibinfo{year}{2025}\natexlab{b}.
\newblock \showarticletitle{Embedding-informed adaptive retrieval-augmented generation of large language models}. In \bibinfo{booktitle}{\emph{Proceedings of the 31st International Conference on Computational Linguistics}}. \bibinfo{pages}{1403--1412}.
\newblock


\bibitem[Huang et~al\mbox{.}(2024)]%
        {huang2024foundation}
\bibfield{author}{\bibinfo{person}{Chengkai Huang}, \bibinfo{person}{Tong Yu}, \bibinfo{person}{Kaige Xie}, \bibinfo{person}{Shuai Zhang}, \bibinfo{person}{Lina Yao}, {and} \bibinfo{person}{Julian McAuley}.} \bibinfo{year}{2024}\natexlab{}.
\newblock \showarticletitle{Foundation models for recommender systems: A survey and new perspectives}.
\newblock \bibinfo{journal}{\emph{arXiv preprint arXiv:2402.11143}} (\bibinfo{year}{2024}).
\newblock


\bibitem[Huang et~al\mbox{.}(2025a)]%
        {huang2025listwise}
\bibfield{author}{\bibinfo{person}{Hongtao Huang}, \bibinfo{person}{Chengkai Huang}, \bibinfo{person}{Junda Wu}, \bibinfo{person}{Tong Yu}, \bibinfo{person}{Julian McAuley}, {and} \bibinfo{person}{Lina Yao}.} \bibinfo{year}{2025}\natexlab{a}.
\newblock \showarticletitle{Listwise Preference Diffusion Optimization for User Behavior Trajectories Prediction}.
\newblock \bibinfo{journal}{\emph{arXiv preprint arXiv:2511.00530}} (\bibinfo{year}{2025}).
\newblock


\bibitem[Jeong et~al\mbox{.}(2024)]%
        {DBLP:conf/naacl/jeong2024adaptive}
\bibfield{author}{\bibinfo{person}{Soyeong Jeong}, \bibinfo{person}{Jinheon Baek}, \bibinfo{person}{Sukmin Cho}, \bibinfo{person}{Sung~Ju Hwang}, {and} \bibinfo{person}{Jong Park}.} \bibinfo{year}{2024}\natexlab{}.
\newblock \showarticletitle{Adaptive-RAG: Learning to Adapt Retrieval-Augmented Large Language Models through Question Complexity}. In \bibinfo{booktitle}{\emph{Proceedings of the 2024 Conference of the North American Chapter of the Association for Computational Linguistics: Human Language Technologies (Volume 1: Long Papers), {NAACL} 2024, Mexico City, Mexico, June 16-21, 2024}}. \bibinfo{publisher}{Association for Computational Linguistics}, \bibinfo{pages}{7036--7050}.
\newblock


\bibitem[Jiang et~al\mbox{.}(2025)]%
        {jiang-etal-2025-rag}
\bibfield{author}{\bibinfo{person}{Jinhao Jiang}, \bibinfo{person}{Jiayi Chen}, \bibinfo{person}{Junyi Li}, \bibinfo{person}{Ruiyang Ren}, \bibinfo{person}{Shijie Wang}, \bibinfo{person}{Xin Zhao}, \bibinfo{person}{Yang Song}, {and} \bibinfo{person}{Tao Zhang}.} \bibinfo{year}{2025}\natexlab{}.
\newblock \showarticletitle{{RAG}-Star: Enhancing Deliberative Reasoning with Retrieval Augmented Verification and Refinement}. In \bibinfo{booktitle}{\emph{Proceedings of the 2025 Conference of the Nations of the Americas Chapter of the Association for Computational Linguistics: Human Language Technologies (Volume 1: Long Papers)}}. \bibinfo{publisher}{Association for Computational Linguistics}, \bibinfo{address}{Albuquerque, New Mexico}, \bibinfo{pages}{7064--7074}.
\newblock
\showISBNx{979-8-89176-189-6}


\bibitem[Johnson et~al\mbox{.}(2019)]%
        {johnson2019billion}
\bibfield{author}{\bibinfo{person}{Jeff Johnson}, \bibinfo{person}{Matthijs Douze}, {and} \bibinfo{person}{Herv{\'e} J{\'e}gou}.} \bibinfo{year}{2019}\natexlab{}.
\newblock \showarticletitle{Billion-scale similarity search with {GPUs}}.
\newblock \bibinfo{journal}{\emph{IEEE Transactions on Big Data}} \bibinfo{volume}{7}, \bibinfo{number}{3} (\bibinfo{year}{2019}), \bibinfo{pages}{535--547}.
\newblock


\bibitem[Joshi et~al\mbox{.}(2017)]%
        {DBLP:conf/acl/Joshi2017tiviaqa}
\bibfield{author}{\bibinfo{person}{Mandar Joshi}, \bibinfo{person}{Eunsol Choi}, \bibinfo{person}{Daniel~S. Weld}, {and} \bibinfo{person}{Luke Zettlemoyer}.} \bibinfo{year}{2017}\natexlab{}.
\newblock \showarticletitle{TriviaQA: {A} Large Scale Distantly Supervised Challenge Dataset for Reading Comprehension}. In \bibinfo{booktitle}{\emph{Proceedings of the 55th Annual Meeting of the Association for Computational Linguistics, {ACL} 2017, Vancouver, Canada, July 30 - August 4, Volume 1: Long Papers}}. \bibinfo{publisher}{Association for Computational Linguistics}, \bibinfo{pages}{1601--1611}.
\newblock


\bibitem[Karpukhin et~al\mbox{.}(2020)]%
        {karpukhin2020dense}
\bibfield{author}{\bibinfo{person}{Vladimir Karpukhin}, \bibinfo{person}{Barlas Oguz}, \bibinfo{person}{Sewon Min}, \bibinfo{person}{Patrick~SH Lewis}, \bibinfo{person}{Ledell Wu}, \bibinfo{person}{Sergey Edunov}, \bibinfo{person}{Danqi Chen}, {and} \bibinfo{person}{Wen-tau Yih}.} \bibinfo{year}{2020}\natexlab{}.
\newblock \showarticletitle{Dense Passage Retrieval for Open-Domain Question Answering.}. In \bibinfo{booktitle}{\emph{EMNLP (1)}}. \bibinfo{pages}{6769--6781}.
\newblock


\bibitem[Kwiatkowski et~al\mbox{.}(2019)]%
        {DBLP:journals/tacl/Kwiatkowski2019nq}
\bibfield{author}{\bibinfo{person}{Tom Kwiatkowski}, \bibinfo{person}{Jennimaria Palomaki}, \bibinfo{person}{Olivia Redfield}, \bibinfo{person}{Michael Collins}, \bibinfo{person}{Ankur~P. Parikh}, \bibinfo{person}{Chris Alberti}, \bibinfo{person}{Danielle Epstein}, \bibinfo{person}{Illia Polosukhin}, \bibinfo{person}{Jacob Devlin}, \bibinfo{person}{Kenton Lee}, \bibinfo{person}{Kristina Toutanova}, \bibinfo{person}{Llion Jones}, \bibinfo{person}{Matthew Kelcey}, \bibinfo{person}{Ming{-}Wei Chang}, \bibinfo{person}{Andrew~M. Dai}, \bibinfo{person}{Jakob Uszkoreit}, \bibinfo{person}{Quoc Le}, {and} \bibinfo{person}{Slav Petrov}.} \bibinfo{year}{2019}\natexlab{}.
\newblock \showarticletitle{Natural Questions: a Benchmark for Question Answering Research}.
\newblock \bibinfo{journal}{\emph{Trans. Assoc. Comput. Linguistics}}  \bibinfo{volume}{7} (\bibinfo{year}{2019}), \bibinfo{pages}{452--466}.
\newblock


\bibitem[Kwon et~al\mbox{.}(2023)]%
        {DBLP:conf/sosp/vllm_kwon}
\bibfield{author}{\bibinfo{person}{Woosuk Kwon}, \bibinfo{person}{Zhuohan Li}, \bibinfo{person}{Siyuan Zhuang}, \bibinfo{person}{Ying Sheng}, \bibinfo{person}{Lianmin Zheng}, \bibinfo{person}{Cody~Hao Yu}, \bibinfo{person}{Joseph Gonzalez}, \bibinfo{person}{Hao Zhang}, {and} \bibinfo{person}{Ion Stoica}.} \bibinfo{year}{2023}\natexlab{}.
\newblock \showarticletitle{Efficient Memory Management for Large Language Model Serving with PagedAttention}. In \bibinfo{booktitle}{\emph{{SOSP}}}. \bibinfo{publisher}{{ACM}}, \bibinfo{pages}{611--626}.
\newblock


\bibitem[Lewis et~al\mbox{.}(2020)]%
        {lewis2020retrieval}
\bibfield{author}{\bibinfo{person}{Patrick Lewis}, \bibinfo{person}{Ethan Perez}, \bibinfo{person}{Aleksandra Piktus}, \bibinfo{person}{Fabio Petroni}, \bibinfo{person}{Vladimir Karpukhin}, \bibinfo{person}{Naman Goyal}, \bibinfo{person}{Heinrich K{\"u}ttler}, \bibinfo{person}{Mike Lewis}, \bibinfo{person}{Wen-tau Yih}, \bibinfo{person}{Tim Rockt{\"a}schel}, {et~al\mbox{.}}} \bibinfo{year}{2020}\natexlab{}.
\newblock \showarticletitle{Retrieval-augmented generation for knowledge-intensive nlp tasks}.
\newblock \bibinfo{journal}{\emph{Advances in neural information processing systems}}  \bibinfo{volume}{33} (\bibinfo{year}{2020}), \bibinfo{pages}{9459--9474}.
\newblock


\bibitem[Li et~al\mbox{.}(2025)]%
        {li2025search}
\bibfield{author}{\bibinfo{person}{Xiaoxi Li}, \bibinfo{person}{Guanting Dong}, \bibinfo{person}{Jiajie Jin}, \bibinfo{person}{Yuyao Zhang}, \bibinfo{person}{Yujia Zhou}, \bibinfo{person}{Yutao Zhu}, \bibinfo{person}{Peitian Zhang}, {and} \bibinfo{person}{Zhicheng Dou}.} \bibinfo{year}{2025}\natexlab{}.
\newblock \showarticletitle{Search-o1: Agentic search-enhanced large reasoning models}.
\newblock \bibinfo{journal}{\emph{arXiv preprint arXiv:2501.05366}} (\bibinfo{year}{2025}).
\newblock


\bibitem[Li et~al\mbox{.}(2021)]%
        {li2021self}
\bibfield{author}{\bibinfo{person}{Yifan Li}, \bibinfo{person}{Xuan Wang}, \bibinfo{person}{Shuhan Qi}, \bibinfo{person}{Chengkai Huang}, \bibinfo{person}{Zoe~L Jiang}, \bibinfo{person}{Qing Liao}, \bibinfo{person}{Jian Guan}, {and} \bibinfo{person}{Jiajia Zhang}.} \bibinfo{year}{2021}\natexlab{}.
\newblock \showarticletitle{Self-supervised learning-based weight adaptive hashing for fast cross-modal retrieval}.
\newblock \bibinfo{journal}{\emph{Signal, Image and Video Processing}} \bibinfo{volume}{15}, \bibinfo{number}{4} (\bibinfo{year}{2021}), \bibinfo{pages}{673--680}.
\newblock


\bibitem[Liu et~al\mbox{.}(2023)]%
        {liu2023lost}
\bibfield{author}{\bibinfo{person}{Nelson~F. Liu}, \bibinfo{person}{Kevin Lin}, \bibinfo{person}{John Hewitt}, \bibinfo{person}{Ashwin Paranjape}, \bibinfo{person}{Michele Bevilacqua}, \bibinfo{person}{Fabio Petroni}, {and} \bibinfo{person}{Percy Liang}.} \bibinfo{year}{2023}\natexlab{}.
\newblock \bibinfo{title}{Lost in the Middle: How Language Models Use Long Contexts}.
\newblock
\showeprint[arxiv]{2307.03172}~[cs.CL]


\bibitem[Meng et~al\mbox{.}(2025)]%
        {DBLP:journals/corr/ultraled}
\bibfield{author}{\bibinfo{person}{Yuang Meng}, \bibinfo{person}{Xin Jin}, \bibinfo{person}{Lina Lei}, \bibinfo{person}{Chun{-}Le Guo}, {and} \bibinfo{person}{Chongyi Li}.} \bibinfo{year}{2025}\natexlab{}.
\newblock \showarticletitle{UltraLED: Learning to See Everything in Ultra-High Dynamic Range Scenes}.
\newblock \bibinfo{journal}{\emph{CoRR}}  \bibinfo{volume}{abs/2510.07741} (\bibinfo{year}{2025}).
\newblock


\bibitem[Peng and Wei(2025)]%
        {DBLP:conf/acl/peng2025grat}
\bibfield{author}{\bibinfo{person}{Xianshu Peng} {and} \bibinfo{person}{Wei Wei}.} \bibinfo{year}{2025}\natexlab{}.
\newblock \showarticletitle{{GRAT:} Guiding Retrieval-Augmented Reasoning through Process Rewards Tree Search}. In \bibinfo{booktitle}{\emph{Proceedings of the 63rd Annual Meeting of the Association for Computational Linguistics (Volume 1: Long Papers), {ACL} 2025, Vienna, Austria, July 27 - August 1, 2025}}. \bibinfo{publisher}{Association for Computational Linguistics}, \bibinfo{pages}{27861--27875}.
\newblock


\bibitem[Press et~al\mbox{.}(2023)]%
        {DBLP:conf/emnlp/PressZMSSL23}
\bibfield{author}{\bibinfo{person}{Ofir Press}, \bibinfo{person}{Muru Zhang}, \bibinfo{person}{Sewon Min}, \bibinfo{person}{Ludwig Schmidt}, \bibinfo{person}{Noah~A. Smith}, {and} \bibinfo{person}{Mike Lewis}.} \bibinfo{year}{2023}\natexlab{}.
\newblock \showarticletitle{Measuring and Narrowing the Compositionality Gap in Language Models}. In \bibinfo{booktitle}{\emph{Findings of the Association for Computational Linguistics: {EMNLP} 2023, Singapore, December 6-10, 2023}}. \bibinfo{publisher}{Association for Computational Linguistics}, \bibinfo{pages}{5687--5711}.
\newblock


\bibitem[Qian et~al\mbox{.}(2025)]%
        {DBLP:conf/www/memorag_qian}
\bibfield{author}{\bibinfo{person}{Hongjin Qian}, \bibinfo{person}{Zheng Liu}, \bibinfo{person}{Peitian Zhang}, \bibinfo{person}{Kelong Mao}, \bibinfo{person}{Defu Lian}, \bibinfo{person}{Zhicheng Dou}, {and} \bibinfo{person}{Tiejun Huang}.} \bibinfo{year}{2025}\natexlab{}.
\newblock \showarticletitle{MemoRAG: Boosting Long Context Processing with Global Memory-Enhanced Retrieval Augmentation}. In \bibinfo{booktitle}{\emph{{WWW}}}. \bibinfo{publisher}{{ACM}}, \bibinfo{pages}{2366--2377}.
\newblock


\bibitem[Rein et~al\mbox{.}(2023)]%
        {DBLP:journals/corr/gpqa_rein}
\bibfield{author}{\bibinfo{person}{David Rein}, \bibinfo{person}{Betty~Li Hou}, \bibinfo{person}{Asa~Cooper Stickland}, \bibinfo{person}{Jackson Petty}, \bibinfo{person}{Richard~Yuanzhe Pang}, \bibinfo{person}{Julien Dirani}, \bibinfo{person}{Julian Michael}, {and} \bibinfo{person}{Samuel~R. Bowman}.} \bibinfo{year}{2023}\natexlab{}.
\newblock \showarticletitle{{GPQA:} {A} Graduate-Level Google-Proof Q{\&}A Benchmark}.
\newblock \bibinfo{journal}{\emph{CoRR}}  \bibinfo{volume}{abs/2311.12022} (\bibinfo{year}{2023}).
\newblock


\bibitem[Roy et~al\mbox{.}(2024)]%
        {roy-etal-2024-contregen}
\bibfield{author}{\bibinfo{person}{Kashob~Kumar Roy}, \bibinfo{person}{Pritom~Saha Akash}, \bibinfo{person}{Kevin Chen-Chuan Chang}, {and} \bibinfo{person}{Lucian Popa}.} \bibinfo{year}{2024}\natexlab{}.
\newblock \showarticletitle{{C}on{TR}e{G}en: Context-driven Tree-structured Retrieval for Open-domain Long-form Text Generation}. In \bibinfo{booktitle}{\emph{Findings of the Association for Computational Linguistics: EMNLP 2024}}. \bibinfo{publisher}{Association for Computational Linguistics}, \bibinfo{address}{Miami, Florida, USA}, \bibinfo{pages}{13773--13784}.
\newblock


\bibitem[Salemi and Zamani(2024)]%
        {DBLP:conf/sigir/salemi2024evaluating}
\bibfield{author}{\bibinfo{person}{Alireza Salemi} {and} \bibinfo{person}{Hamed Zamani}.} \bibinfo{year}{2024}\natexlab{}.
\newblock \showarticletitle{Evaluating Retrieval Quality in Retrieval-Augmented Generation}. In \bibinfo{booktitle}{\emph{Proceedings of the 47th International {ACM} {SIGIR} Conference on Research and Development in Information Retrieval, {SIGIR} 2024, Washington DC, USA, July 14-18, 2024}}. \bibinfo{publisher}{{ACM}}, \bibinfo{pages}{2395--2400}.
\newblock


\bibitem[Shen et~al\mbox{.}(2024)]%
        {shen2024understanding}
\bibfield{author}{\bibinfo{person}{Michael Shen}, \bibinfo{person}{Muhammad Umar}, \bibinfo{person}{Kiwan Maeng}, \bibinfo{person}{G.~Edward Suh}, {and} \bibinfo{person}{Udit Gupta}.} \bibinfo{year}{2024}\natexlab{}.
\newblock \bibinfo{title}{Towards Understanding Systems Trade-offs in Retrieval-Augmented Generation Model Inference}.
\newblock
\showeprint[arxiv]{2412.11854}~[cs.AR]


\bibitem[Shen et~al\mbox{.}(2025)]%
        {DBLP:conf/coling/shen2025reasoningtree}
\bibfield{author}{\bibinfo{person}{Tiesunlong Shen}, \bibinfo{person}{Jin Wang}, \bibinfo{person}{Xuejie Zhang}, {and} \bibinfo{person}{Erik Cambria}.} \bibinfo{year}{2025}\natexlab{}.
\newblock \showarticletitle{Reasoning with Trees: Faithful Question Answering over Knowledge Graph}. In \bibinfo{booktitle}{\emph{Proceedings of the 31st International Conference on Computational Linguistics, {COLING} 2025, Abu Dhabi, UAE, January 19-24, 2025}}. \bibinfo{publisher}{Association for Computational Linguistics}, \bibinfo{pages}{3138--3157}.
\newblock


\bibitem[Shuster et~al\mbox{.}(2021)]%
        {DBLP:conf/emnlp/shuster2021retrieval}
\bibfield{author}{\bibinfo{person}{Kurt Shuster}, \bibinfo{person}{Spencer Poff}, \bibinfo{person}{Moya Chen}, \bibinfo{person}{Douwe Kiela}, {and} \bibinfo{person}{Jason Weston}.} \bibinfo{year}{2021}\natexlab{}.
\newblock \showarticletitle{Retrieval Augmentation Reduces Hallucination in Conversation}. In \bibinfo{booktitle}{\emph{Findings of the Association for Computational Linguistics: {EMNLP} 2021, Virtual Event / Punta Cana, Dominican Republic, 16-20 November, 2021}}. \bibinfo{publisher}{Association for Computational Linguistics}, \bibinfo{pages}{3784--3803}.
\newblock


\bibitem[Singh et~al\mbox{.}(2025)]%
        {singh2025chunkrag}
\bibfield{author}{\bibinfo{person}{Ishneet~Sukhvinder Singh}, \bibinfo{person}{Ritvik Aggarwal}, \bibinfo{person}{Ibrahim Allahverdiyev}, \bibinfo{person}{Muhammad Taha}, \bibinfo{person}{Aslihan Akalin}, \bibinfo{person}{Kevin Zhu}, {and} \bibinfo{person}{Sean O'Brien}.} \bibinfo{year}{2025}\natexlab{}.
\newblock \bibinfo{title}{ChunkRAG: Novel LLM-Chunk Filtering Method for RAG Systems}.
\newblock
\showeprint[arxiv]{2410.19572}~[cs.CL]


\bibitem[Su et~al\mbox{.}(2024)]%
        {DBLP:conf/acl/su2024dragin}
\bibfield{author}{\bibinfo{person}{Weihang Su}, \bibinfo{person}{Yichen Tang}, \bibinfo{person}{Qingyao Ai}, \bibinfo{person}{Zhijing Wu}, {and} \bibinfo{person}{Yiqun Liu}.} \bibinfo{year}{2024}\natexlab{}.
\newblock \showarticletitle{{DRAGIN:} Dynamic Retrieval Augmented Generation based on the Real-time Information Needs of Large Language Models}. In \bibinfo{booktitle}{\emph{Proceedings of the 62nd Annual Meeting of the Association for Computational Linguistics (Volume 1: Long Papers), {ACL} 2024, Bangkok, Thailand, August 11-16, 2024}}. \bibinfo{publisher}{Association for Computational Linguistics}, \bibinfo{pages}{12991--13013}.
\newblock


\bibitem[Sun et~al\mbox{.}(2025)]%
        {DBLP:conf/acl/sun2025era}
\bibfield{author}{\bibinfo{person}{Hao Sun}, \bibinfo{person}{Hengyi Cai}, \bibinfo{person}{Yuchen Li}, \bibinfo{person}{Xuanbo Fan}, \bibinfo{person}{Xiaochi Wei}, \bibinfo{person}{Shuaiqiang Wang}, \bibinfo{person}{Yan Zhang}, {and} \bibinfo{person}{Dawei Yin}.} \bibinfo{year}{2025}\natexlab{}.
\newblock \showarticletitle{Enhancing Retrieval-Augmented Generation via Evidence Tree Search}. In \bibinfo{booktitle}{\emph{Proceedings of the 63rd Annual Meeting of the Association for Computational Linguistics (Volume 1: Long Papers), {ACL} 2025, Vienna, Austria, July 27 - August 1, 2025}}. \bibinfo{publisher}{Association for Computational Linguistics}, \bibinfo{pages}{24116--24127}.
\newblock


\bibitem[Touvron et~al\mbox{.}(2023)]%
        {DBLP:journals/corr/llama2}
\bibfield{author}{\bibinfo{person}{Hugo Touvron}, \bibinfo{person}{Louis Martin}, \bibinfo{person}{Kevin Stone}, {and} \bibinfo{person}{et al.}} \bibinfo{year}{2023}\natexlab{}.
\newblock \showarticletitle{Llama 2: Open Foundation and Fine-Tuned Chat Models}.
\newblock \bibinfo{journal}{\emph{CoRR}}  \bibinfo{volume}{abs/2307.09288} (\bibinfo{year}{2023}).
\newblock


\bibitem[Trivedi et~al\mbox{.}(2022a)]%
        {trivedi2022ircot}
\bibfield{author}{\bibinfo{person}{Harsh Trivedi}, \bibinfo{person}{Niranjan Balasubramanian}, \bibinfo{person}{Tushar Khot}, {and} \bibinfo{person}{Ashish Sabharwal}.} \bibinfo{year}{2022}\natexlab{a}.
\newblock \showarticletitle{Interleaving retrieval with chain-of-thought reasoning for knowledge-intensive multi-step questions}.
\newblock \bibinfo{journal}{\emph{arXiv preprint arXiv:2212.10509}} (\bibinfo{year}{2022}).
\newblock


\bibitem[Trivedi et~al\mbox{.}(2022b)]%
        {DBLP:journals/tacl/TrivediBKS22}
\bibfield{author}{\bibinfo{person}{Harsh Trivedi}, \bibinfo{person}{Niranjan Balasubramanian}, \bibinfo{person}{Tushar Khot}, {and} \bibinfo{person}{Ashish Sabharwal}.} \bibinfo{year}{2022}\natexlab{b}.
\newblock \showarticletitle{MuSiQue: Multihop Questions via Single-hop Question Composition}.
\newblock \bibinfo{journal}{\emph{Trans. Assoc. Comput. Linguistics}}  \bibinfo{volume}{10} (\bibinfo{year}{2022}), \bibinfo{pages}{539--554}.
\newblock


\bibitem[Wang et~al\mbox{.}(2022)]%
        {DBLP:journals/corr/wang2022e5}
\bibfield{author}{\bibinfo{person}{Liang Wang}, \bibinfo{person}{Nan Yang}, \bibinfo{person}{Xiaolong Huang}, {and} \bibinfo{person}{et al.}} \bibinfo{year}{2022}\natexlab{}.
\newblock \showarticletitle{Text Embeddings by Weakly-Supervised Contrastive Pre-training}.
\newblock \bibinfo{journal}{\emph{CoRR}}  \bibinfo{volume}{abs/2212.03533} (\bibinfo{year}{2022}).
\newblock
\showeprint[arXiv]{2212.03533}


\bibitem[Wang et~al\mbox{.}(2024)]%
        {wang2024forgetting}
\bibfield{author}{\bibinfo{person}{Shang Wang}, \bibinfo{person}{Tianqing Zhu}, \bibinfo{person}{Dayong Ye}, {and} \bibinfo{person}{Wanlei Zhou}.} \bibinfo{year}{2024}\natexlab{}.
\newblock \bibinfo{title}{When Machine Unlearning Meets Retrieval-Augmented Generation (RAG): Keep Secret or Forget Knowledge?}
\newblock
\showeprint[arxiv]{2410.15267}~[cs.CR]


\bibitem[Wei et~al\mbox{.}(2022)]%
        {wei2022chain}
\bibfield{author}{\bibinfo{person}{Jason Wei}, \bibinfo{person}{Xuezhi Wang}, \bibinfo{person}{Dale Schuurmans}, \bibinfo{person}{Maarten Bosma}, \bibinfo{person}{Fei Xia}, \bibinfo{person}{Ed Chi}, \bibinfo{person}{Quoc~V Le}, \bibinfo{person}{Denny Zhou}, {et~al\mbox{.}}} \bibinfo{year}{2022}\natexlab{}.
\newblock \showarticletitle{Chain-of-thought prompting elicits reasoning in large language models}.
\newblock \bibinfo{journal}{\emph{Advances in neural information processing systems}}  \bibinfo{volume}{35} (\bibinfo{year}{2022}), \bibinfo{pages}{24824--24837}.
\newblock


\bibitem[Xiao et~al\mbox{.}(2023)]%
        {DBLP:journals/corr/xiao2023bge}
\bibfield{author}{\bibinfo{person}{Shitao Xiao}, \bibinfo{person}{Zheng Liu}, \bibinfo{person}{Peitian Zhang}, {and} \bibinfo{person}{Niklas Muennighoff}.} \bibinfo{year}{2023}\natexlab{}.
\newblock \showarticletitle{C-Pack: Packaged Resources To Advance General Chinese Embedding}.
\newblock \bibinfo{journal}{\emph{CoRR}}  \bibinfo{volume}{abs/2309.07597} (\bibinfo{year}{2023}).
\newblock
\showeprint[arXiv]{2309.07597}


\bibitem[Xu et~al\mbox{.}(2024)]%
        {xu2024search}
\bibfield{author}{\bibinfo{person}{Shicheng Xu}, \bibinfo{person}{Liang Pang}, \bibinfo{person}{Huawei Shen}, \bibinfo{person}{Xueqi Cheng}, {and} \bibinfo{person}{Tat-Seng Chua}.} \bibinfo{year}{2024}\natexlab{}.
\newblock \showarticletitle{Search-in-the-chain: Interactively enhancing large language models with search for knowledge-intensive tasks}. In \bibinfo{booktitle}{\emph{Proceedings of the ACM Web Conference 2024}}. \bibinfo{pages}{1362--1373}.
\newblock


\bibitem[Yang et~al\mbox{.}(2025)]%
        {yang2025qwen3technicalreport}
\bibfield{author}{\bibinfo{person}{An Yang}, \bibinfo{person}{Anfeng Li}, \bibinfo{person}{Baosong Yang}, {and} \bibinfo{person}{et al.}} \bibinfo{year}{2025}\natexlab{}.
\newblock \bibinfo{title}{Qwen3 Technical Report}.
\newblock
\showeprint[arxiv]{2505.09388}~[cs.CL]


\bibitem[Yang et~al\mbox{.}(2018)]%
        {DBLP:conf/emnlp/Yang0ZBCSM18}
\bibfield{author}{\bibinfo{person}{Zhilin Yang}, \bibinfo{person}{Peng Qi}, \bibinfo{person}{Saizheng Zhang}, \bibinfo{person}{Yoshua Bengio}, \bibinfo{person}{William~W. Cohen}, \bibinfo{person}{Ruslan Salakhutdinov}, {and} \bibinfo{person}{Christopher~D. Manning}.} \bibinfo{year}{2018}\natexlab{}.
\newblock \showarticletitle{HotpotQA: {A} Dataset for Diverse, Explainable Multi-hop Question Answering}. In \bibinfo{booktitle}{\emph{Proceedings of the 2018 Conference on Empirical Methods in Natural Language Processing, Brussels, Belgium, October 31 - November 4, 2018}}. \bibinfo{publisher}{Association for Computational Linguistics}, \bibinfo{pages}{2369--2380}.
\newblock


\bibitem[Yao et~al\mbox{.}(2023a)]%
        {yao2023tree}
\bibfield{author}{\bibinfo{person}{Shunyu Yao}, \bibinfo{person}{Dian Yu}, \bibinfo{person}{Jeffrey Zhao}, \bibinfo{person}{Izhak Shafran}, \bibinfo{person}{Tom Griffiths}, \bibinfo{person}{Yuan Cao}, {and} \bibinfo{person}{Karthik Narasimhan}.} \bibinfo{year}{2023}\natexlab{a}.
\newblock \showarticletitle{Tree of thoughts: Deliberate problem solving with large language models}.
\newblock \bibinfo{journal}{\emph{Advances in neural information processing systems}}  \bibinfo{volume}{36} (\bibinfo{year}{2023}), \bibinfo{pages}{11809--11822}.
\newblock


\bibitem[Yao et~al\mbox{.}(2023b)]%
        {yao2023react}
\bibfield{author}{\bibinfo{person}{Shunyu Yao}, \bibinfo{person}{Jeffrey Zhao}, \bibinfo{person}{Dian Yu}, \bibinfo{person}{Nan Du}, \bibinfo{person}{Izhak Shafran}, \bibinfo{person}{Karthik Narasimhan}, {and} \bibinfo{person}{Yuan Cao}.} \bibinfo{year}{2023}\natexlab{b}.
\newblock \showarticletitle{React: Synergizing reasoning and acting in language models}. In \bibinfo{booktitle}{\emph{International Conference on Learning Representations (ICLR)}}.
\newblock


\end{thebibliography}

\appendix

\section{Appendix}

\subsection{Datasets}
\label{app:datasets}
We evaluate our method on six QA datasets of varying complexity. Among them, four datasets feature complex, multi-hop reasoning challenges such as cross-passage evidence integration, factual alignment, entity linking, and logical composition:

\begin{itemize}[leftmargin=1em, itemindent=0em, itemsep=0.3em]
    \item \textbf{HotpotQA} \cite{DBLP:conf/emnlp/Yang0ZBCSM18}: A widely used multi-hop QA benchmark where each question requires integrating evidence from multiple documents. We follow the full wiki setting. 
    \item \textbf{2WikiMultihopQA} \cite{DBLP:conf/coling/HoNSA20}: Constructed from Wikipedia with explicitly structured multi-hop questions, making it well-suited for evaluating path-based and compositional reasoning. 
    \item \textbf{MuSiQue-full} \cite{DBLP:journals/tacl/TrivediBKS22}: Composed of multiple independently meaningful sub-questions, naturally aligned with query-tree style decomposition and testing information integration ability. 
    \item \textbf{BAMBOOGLE} \cite{DBLP:conf/emnlp/PressZMSSL23}: A challenging dataset emphasizing instruction-style QA, with frequent intent shifts and cross-sentence constraints that demand structured reasoning. 
    \item \textbf{GPQA} \cite{DBLP:journals/corr/gpqa_rein}: A large-scale benchmark of graduate-level problem-solving questions, designed to stress-test compositional reasoning and domain knowledge coverage. 
\end{itemize}

To verify the adaptive capability of \ModelName{} across varying levels of reasoning complexity, we include two single-hop QA datasets: 

\begin{itemize}[leftmargin=1em, itemindent=0em, itemsep=0.3em]
    \item \textbf{Natural Questions (NQ)} \cite{DBLP:journals/tacl/Kwiatkowski2019nq}: Real search queries paired with Wikipedia passages, with most questions answerable from a single paragraph, serving as a factual QA baseline. 
    \item \textbf{TriviaQA} \cite{DBLP:conf/acl/Joshi2017tiviaqa}: Trivia-style questions with short entity answers and minimal reasoning, suitable for evaluating fact retrieval.
\end{itemize}

Following \citet{jiang-etal-2025-rag}, we randomly sample 500 validation instances as the test set for each dataset, except for BAMBOOGLE, where all 125 available instances are used.  
Notably, only HotpotQA, 2WikiMultihopQA, and MuSiQue provide annotated golden documents, enabling the computation of Evidence Forgetting Rate (EFR). For the remaining datasets, evaluation focuses on accuracy and efficiency metrics, ensuring comprehensive assessment under both golden-evidence and open-domain settings.

\subsection{Implementation Details}
\label{app:details}
In all experiments, we adopt large language models (LLMs) as the backbone generators and evaluate the performance of various methods under different retrieval strategies. Specifically, we select two mainstream open-source models as base models: \textbf{LLaMA3.1-8B-Instruct} \cite{grattafiori2024llama3herdmodels} and \textbf{Qwen3-8B} \cite{yang2025qwen3technicalreport}. Since the Self-RAG method is fine-tuned on the \texttt{Llama2-7b-HF} \cite{DBLP:journals/corr/llama2} model, we directly use the released checkpoint provided by the original authors to ensure fairness and comparability. To further assess the generalizability of our method across models of different parameter scales, we additionally evaluate on two larger models: \textbf{LLaMA3.1-70B-Instruct} and \textbf{Qwen3-32B}.

For the retriever component, we evaluate two widely used dense retrieval strategies: \textbf{BGE} \cite{DBLP:journals/corr/xiao2023bge} and \textbf{E5} \cite{DBLP:journals/corr/wang2022e5}. In the main experiments, following the RAG-Star setup, we use \textbf{BGE-large-en-v1.5} \cite{DBLP:journals/corr/xiao2023bge} as the default retriever. Each sub-query retrieves the top-5 documents, which are used to construct the retrieval-augmented context. For the knowledge corpus, we adopt the 2018 English Wikipedia dump used in DPR \cite{karpukhin2020dense}, and build dense indexes through a unified preprocessing pipeline to ensure consistency across retrievers. Results based on E5 are included in the Appendix \ref{app:e5}.

For \ModelName{}, we set the tree branching factor to 2, maximum depth to 3, and the confidence threshold $\tau_A$ to 0.95. All hyperparameter settings for baseline methods follow their original implementations. During inference, we set \texttt{max\_tokens} to 4096 for all models. For Qwen-based models, we observe frequent token repetition and apply the official remedy by setting \texttt{repetition\_penalty} to 1.05 to mitigate this issue. 

Considering that most baselines adopt multi-step querying strategies, for fair comparison, we cap the number of retriever calls to 7 for all methods to ensure a fair comparison. This limit corresponds to the configuration of \ModelName{} with depth 3, where a full ternary tree contains up to 7 nodes.

All experiments are conducted on a server equipped with \textbf{4$\times$ NVIDIA L40 PCIe 48GB GPUs}, and inference is performed using the \textbf{vLLM} framework \cite{DBLP:journals/corr/ultraled,DBLP:conf/sosp/vllm_kwon}.

\subsection{Evaluation on Higher-Complexity Datasets}

\begin{table}[t]
\centering
\caption{Comparison of answer accuracy (EM/F1) and inference latency across RAG methods on higher-complexity datasets (Bamboogle and GPQA). Bold indicates the best result, and \underline{underline} indicates the second-best. Inference latency is compared only among multi-turn retrieval methods.}

\begin{tabular}{c c c c c c}
\toprule
\multirow{2}{*}{\textbf{Methods}} & 
\multirow{2}{*}{\makecell{\textbf{Time Cost} \\ (ms)}} & 
\multicolumn{2}{c}{\textbf{Bamboogle}} & 
\multicolumn{2}{c}{\textbf{GPQA}} \\
 \cmidrule(lr){3-4} \cmidrule(lr){5-6} 
 & & \textbf{EM} $\uparrow$ & \textbf{F1} $\uparrow$ & \textbf{EM} $\uparrow$ & \textbf{F1} $\uparrow$  \\
\midrule
\rowcolor{gray!20} \multicolumn{6}{c}{\textit{Llama-3.1-8B-Instruct}} \\
 Vanilla & 153 & 12.0 & 19.7 & \underline{24.7} & \underline{24.7} \\
\midrule
 Standard RAG & 813 & 24.0 & 32.5 & 24.2 & 24.2 \\
 MemoRAG & 10,405 & 17.6 & 27.2 & 20.2 & 20.3 \\
\midrule
 React & 4,637 & 14.4 & 18.9 & 0.5 & 0.6 \\
 Search-o1 & \underline{4,024} & 19.2 & 23.7 & 6.0 & 6.3 \\
 Self-RAG & 3,634 & 7.2 & 15.7 & 3.0 & 3.6 \\
 ConTReGen & 6,022 & 24.0 & 31.5 & 14.1 & 14.5  \\
 RAG-Star & 31,253 & 19.2 & 24.9 & 11.1 & 11.5 \\
 ProbTree & 4,035 & \underline{24.8} & \underline{34.1} & 13.6 & 13.7  \\
 \ModelName{} (Ours) & \textbf{855} & \textbf{26.4} & \textbf{38.8} & \textbf{29.7} & \textbf{29.7}  \\
\midrule
\rowcolor{gray!20} \multicolumn{6}{c}{\textit{Qwen3-8B}} \\
 Vanilla & 3,433 & \underline{33.6} & \underline{45.3} & 47.4 & 47.4 \\
\midrule
 Standard RAG & 3,673 & 31.2 & 38.1 & \underline{50.5} & \underline{50.5} \\
 MemoRAG & 33,966 & 17.6 & 27.8 & 35.8 & 35.8 \\
\midrule
 React & 10,872 & 14.4 & 18.4 & 7.5 & 8.2 \\
 Search-o1 & 9,527 & 24.8 & 29.2 & 30.3 & 30.3 \\
 Self-RAG & \textbf{3,634} & 7.2 & 15.7 & 3.0 & 3.6 \\
 ConTReGen & 18,830 & 30.4 & 39.2 & 51.0 & 51.0 \\
 RAG-Star & 12,822 & 32.8 & 44.9 & 47.9 & 47.9 \\
 ProbTree & 9,304 & 16.8 & 22.4 & 37.3 & 37.3 \\
 \ModelName{} (Ours) & \underline{8,895} & \textbf{34.4} & \textbf{46.7} & \textbf{53.5} & \textbf{53.5} \\
\bottomrule
\end{tabular}
\label{tab:supp_main_results}
\end{table}

To further validate robustness, we evaluate \ModelName{} on two more complex datasets, Bamboogle and GPQA. As shown in Table~\ref{tab:supp_main_results}, baseline methods generally show substantial degradation on these tasks, highlighting the challenge of retrieving and reasoning over long-tail knowledge. In contrast, \ModelName{} consistently outperforms all alternatives in EM and F1 under both LLaMA-3.1-8B and Qwen3-8B. For instance, it achieves 26.4\% EM and 38.8\% F1 on Bamboogle with LLaMA-3.1-8B, and 53.5\% EM and 53.5\% F1 on GPQA with Qwen3-8B, surpassing the strongest baselines while maintaining lower inference cost. Since these datasets do not provide golden evidence annotations, the Evidence Forgetting Rate (EFR) cannot be computed. Overall, the results confirm that the mechanisms of confidence-guided pruning and dynamic decomposition remain effective even under higher reasoning complexity.

\subsection{Evaluation with Larger-Scale Base Models}
\label{app:larger_models}
\begin{table}[t]
\centering
\caption{Performance Comparison of \ModelName{} and Baselines on Large-Scale Language Models on HotpotQA}
\setlength{\tabcolsep}{0.8mm}
\begin{tabular}{c c c c c c c c c}
\toprule
\multirow{2}{*}{\textbf{Methods}} & 
\multicolumn{4}{c}{\textbf{Llama-3.1-70B-Instruct}} &
\multicolumn{4}{c}{\textbf{Qwen3-32B}} \\
\cmidrule(lr){2-5} \cmidrule(lr){6-9}
& \textbf{Time} & \textbf{EM} & \textbf{F1}  & \textbf{EFR} 
& \textbf{Time} & \textbf{EM}  & \textbf{F1}  & \textbf{EFR}  \\
\midrule
Vanilla & 515 & 25.0 & 30.8 & - & 3,459 & 21.2 & 25.4 & - \\
\midrule
RAG & 603 & \underline{50.6} & 57.2 & \underline{29.7} & 1,394 & \underline{54.2} & \underline{62.6} & \textbf{21.9} \\
MemoRAG & 1,000 & 10.0 & 10.0 & 10.0 & 1,000 & 10.0 & 10.0 & 10.0 \\
\midrule
React & 2,824 & 41.2 & 46.6 & 41.8 & 7,927 & 34.4 & 40.4 & 35.5  \\
Search-o1 & 3,057 & 46.2 & 52.6 & 40.2 & 5,047 & 42.4 & 49.4 & 43.6 \\
Self-RAG & 5,294 & 30.6 & 37.5 & 59.1 & 5,294 & 30.6 & 37.5 & 59.1 \\
ConTRGen & 22,799 & 50.6 & \underline{57.7} & 30.7 & 16,921 & 49.8 & 58.7 & 36.4 \\
RAG-Star & 43,239 & 41.6 & 50.9 & 49.6 & 52,527 & 38.4 & 46.9 & 49.3 \\
ProbTree & \underline{2,583} & 36.4 & 44.6 & 50.5 & \underline{2,504} & 25.6 & 38.1 & 59.0 \\
\ModelName{} & \textbf{1,432} & \textbf{58.2} & \textbf{66.0} & \textbf{29.2} & \textbf{2,221} & \textbf{58.4} & \textbf{66.0} & \underline{24.6} \\
\bottomrule
\end{tabular}
\label{tab:results_bigger}
\end{table}

\begin{table*}[ht]
\centering
\caption{Performance of RAG methods using the E5 retriever on multi-hop QA datasets. Inference latency and EFR are compared only among multi-turn retrieval methods.}
\renewcommand{\arraystretch}{0.9}
\begin{tabular}{c c c c c c c c c c c c}
\toprule
\multirow{2}{*}{\textbf{Types}} & 
\multirow{2}{*}{\textbf{Methods}} & 
\multirow{2}{*}{\makecell{\textbf{Time Cost} \\ (ms)}} & 
\multicolumn{3}{c}{\textbf{HotpotQA}} & 
\multicolumn{3}{c}{\textbf{2WikiQA}} & 
\multicolumn{3}{c}{\textbf{MusiQue}} \\
 \cmidrule(lr){4-6} \cmidrule(lr){7-9} \cmidrule(lr){10-12} 
& & & \textbf{EM} $\uparrow$ & \textbf{F1} $\uparrow$ & \textbf{EFR} $\downarrow$ & \textbf{EM} $\uparrow$ & \textbf{F1} $\uparrow$ & \textbf{EFR} $\downarrow$ & \textbf{EM} $\uparrow$ & \textbf{F1} $\uparrow$ & \textbf{EFR} $\downarrow$\\
\midrule
\rowcolor{gray!20} \multicolumn{12}{c}{\textit{Llama-3.1-8B-Instruct}} \\
\textit{No-Retrieval} & Vanilla & 64 & 25.6 & 30.8 & - & 18.6 & 25.0 & - & 5.2 & 9.5 & - \\
\midrule
\multirow{2}{*}{\textit{Single-Retrieval}} & Standard RAG & 277 & \underline{44.8} & 51.3 & 30.5 & 17.0 & 22.5 & 35.7 & 7.0 & 11.0 & - \\
& MemoRAG & 8,562 & 31.8 & 39.1 & 58.1 & 19.8 & 25.9 & 74.4 & 8.0 & 11.7 & - \\
\midrule
\multirow{7}{*}{\textit{Multi-Retrieval}} & React & 3,144 & 23.4 & 27.0 & 60.0 & 17.4 & 21.7 & 70.0 & 8.0 & 12.0 & 70.3 \\
& Search-o1 & 2,278 & 33.4 & 37.6 & 55.9 & 22.6 & 27.6 & 64.8 & \underline{12.6} & \underline{16.7} & \underline{58.1} \\
& Self-RAG & 3,107 & 29.0 & 36.0 & 59.2 & 15.6 & 26.4 & 83.3 & 6.8 & 13.0 & 83.3 \\
& ConTReGen & \underline{1,864} & 40.2 & 46.5 & \underline{41.0} & 19.2 & 24.0 & \textbf{42.8} & 7.8 & 10.8 & 80.0 \\
& RAG-Star & 8,265 & 42.0 & \underline{54.4} & 49.7 & \underline{34.0} & \underline{42.0} & 67.5 & 10.2 & 14.7 & 88.4 \\
& ProbTree & 3,718 & 31.0 & 38.3 & 51.7 & 31.6 & 36.6 & 49.3 & 9.4 & 14.5 & 80.0  \\
& \ModelName{} (Ours) & \textbf{746} & \textbf{48.6} & \textbf{56.5} & \textbf{30.5} & \textbf{35.8} & \textbf{42.8} & \underline{46.3} & \textbf{13.8} & \textbf{22.4} & \textbf{55.6} \\
\midrule
\rowcolor{gray!20} \multicolumn{12}{c}{\textit{Qwen3-8B}} \\
\textit{No-Retrieval}& Vanilla & 740 & 27.2 & 36.2 & - & 30.8 & 35.6 & - & 10.2 & 16.2 & - \\
\midrule
\multirow{2}{*}{\textit{Single-Retrieval}} & Standard RAG & 1,221 & \underline{46.6} & \underline{53.8} & 30.5 & 24.0 & 28.3 & 23.8 & 10.8 & 15.6 & - \\
& MemoRAG & 10,669 & 35.6 & 46.5 & 50.0 & 22.6 & 30.9 & 41.8 & 8.6 & 13.9 & - \\
\midrule
\multirow{7}{*}{\textit{Multi-Retrieval}}  & React & 7,192 & 30.0 & 36.2 & 48.7 & 25.0 & 30.0 & 53.5 & 11.2 & 15.8 & \underline{50.0} \\
& Search-o1 & 8,296 & 34.2 & 40.9 & 54.6 & 26.4 & 31.6 & 48.0 & \underline{13.4} & 19.8 & 62.7 \\
& Self-RAG & \underline{3,107} & 29.0 & 36.0 & 59.2 & 15.6 & 26.4 & 83.3 & 6.8 & 13.0 & 83.3 \\
& ConTReGen & 4,897 & 44.8 & 52.5 & \underline{36.0} & 27.4 & 31.6 & \underline{30.3} & 11.6 & 16.7 & \underline{50.0} \\
& RAG-Star & 10,141 & 45.2 & 51.7 & 55.4 & \underline{37.2} & \underline{42.5} & 65.1 & 12.6 & \underline{20.1} & 72.7 \\
& ProbTree & 3,633 & 33.6 & 40.3 & 49.3 & 27.6 & 31.1 & 51.9 & 5.2 & 10.9 & 85.7 \\
& \ModelName{} (Ours) & \textbf{2,849} & \textbf{52.2} & \textbf{59.5} & \textbf{29.9} & \textbf{40.2} & \textbf{46.0} &  
\textbf{27.7} & \textbf{16.0} & \textbf{23.6} & \textbf{40.7} \\
\bottomrule
\end{tabular}
\label{tab:app_e5}
\end{table*}

We further investigate whether increasing the scale of backbone models can inherently alleviate evidence forgetting. Table~\ref{tab:results_bigger} presents results on HotpotQA using LLaMA-3.1-70B-Instruct and Qwen3-32B. Although larger models generally achieve stronger accuracy than their smaller counterparts, the phenomenon of evidence forgetting remains evident: multi-turn baselines such as ReAct, Search-o1, and RAG-Star still exhibit high EFR values (often exceeding 40\%), and even tree-structured methods such as ConTReGen and ProbTree do not fully overcome this limitation.  

In contrast, \ModelName{} consistently achieves the best overall performance across both backbones. On LLaMA-3.1-70B, it reaches 58.2\% EM and 66.0\% F1 while reducing EFR to 29.2\%. Similarly, on Qwen3-32B, it achieves 58.4\% EM and 66.0\% F1 with an EFR of 24.6\%, ranking best or second-best across all metrics. These improvements highlight that confidence-guided pruning and adaptive decomposition remain indispensable, even when the underlying language model has substantially larger capacity.  

Overall, these findings demonstrate that scaling up model size alone does not resolve the challenge of evidence forgetting. Instead, principled mechanisms such as those introduced in \ModelName{} are crucial for ensuring reliable reasoning and maximizing the utility of retrieved evidence in large-scale settings.

\subsection{Evaluation with Alternative Retriever}
\label{app:e5}

To further examine the robustness of our framework, we replace the default retriever with the E5 embedding model and re-evaluate on three representative datasets. As shown in Table~\ref{tab:app_e5}, evidence forgetting remains a prominent issue across baselines: multi-turn methods such as ReAct, Search-o1, and RAG-Star continue to suffer from high EFR values, often above 48\%, despite the use of a different retriever. These results indicate that the forgetting phenomenon is not solely attributable to retriever quality, but persists as a systematic weakness of iterative retrieval paradigms.

In contrast, \ModelName{} consistently delivers superior performance. Under both LLaMA-3.1-8B and Qwen3-8B backbones, it achieves the highest EM and F1 scores on HotpotQA, 2WikiQA, and MusiQue, while also maintaining lower or second-lowest EFR values. For example, on HotpotQA, \ModelName{} attains 52.2\% EM and 59.5\% F1 with an EFR of 29.9\% under Qwen3-8B, outperforming all multi-turn baselines by a clear margin. At the same time, it achieves these improvements with substantially reduced inference latency compared to other multi-turn methods.

Overall, these findings demonstrate that even with a more advanced retriever, evidence forgetting remains an inherent challenge. \ModelName{} effectively mitigates this issue and continues to achieve strong accuracy, low forgetting, and competitive efficiency, confirming the robustness and generalizability of its design.

\subsection{Case Study}\label{app:case}

\begin{figure*}[ht]
\centering
\includegraphics[width=\textwidth]{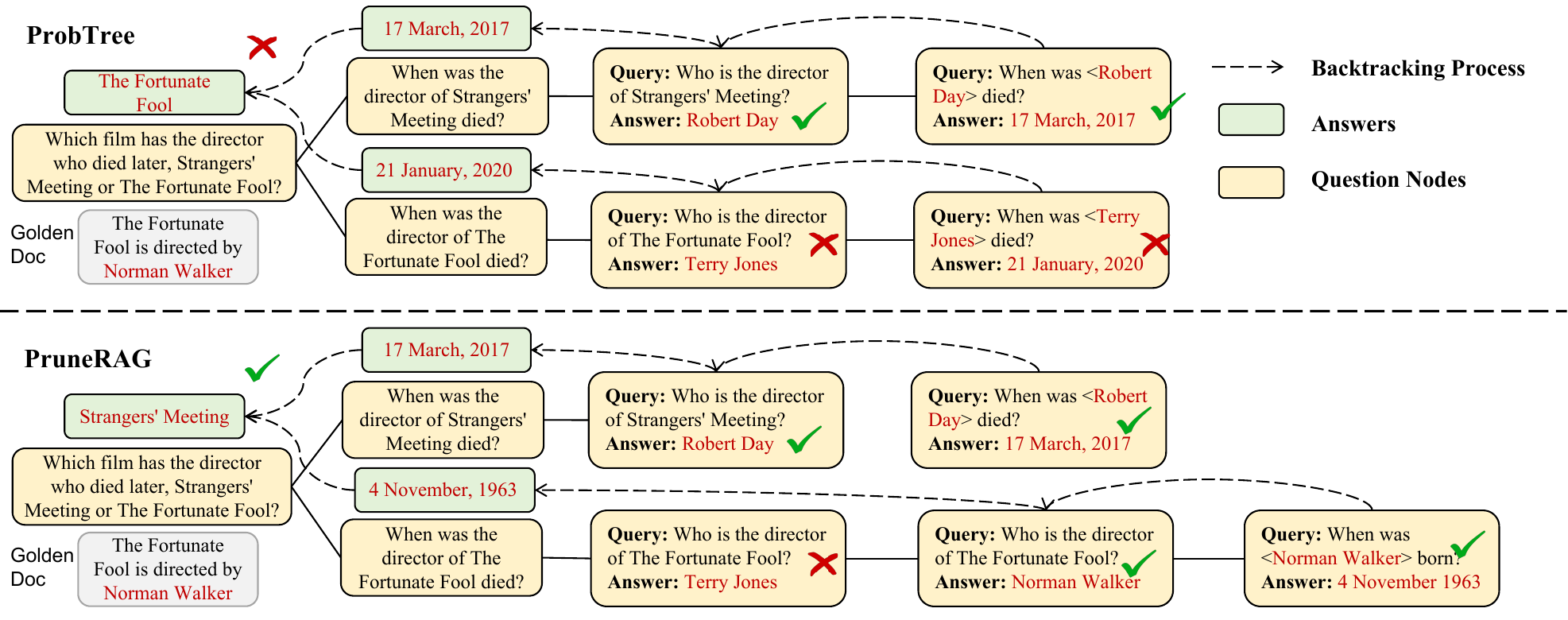} 
\caption{Case study: ProbTree suffers from evidence forgetting, while PruneRAG prunes errors and recovers the correct reasoning chain.}
\label{case_study}
\end{figure*}

Figure~\ref{case_study} illustrates an example of evidence forgetting when comparing the directors of \textit{Strangers’ Meeting} and \textit{The Fortunate Fool}. ProbTree first retrieves the correct evidence that \textit{The Fortunate Fool} was directed by Norman Walker, but incorrectly accepts ''\textbf{Terry Jones}'' as an intermediate answer. Without pruning, this error propagates and the reasoning path diverges, causing the golden evidence to be lost and the final answer to be wrong.

In contrast, \ModelName{} prunes the low-confidence branch and regenerates a sub-query consistent with the retrieved evidence, correctly recovering Norman Walker. This allows the reasoning chain to integrate the true death dates and reach the correct conclusion. The case illustrates how \ModelName{} suppresses low-quality outputs, preserves key evidence, and maintains robust multi-hop reasoning.

\section{Limitations}
Despite its effectiveness, our approach inherits several limitations related to representation quality and long-horizon alignment. First, the method relies on similarity-based retrieval over learned representations, which may be sensitive to representation mismatch under domain shift or noisy supervision. Prior work in cross-modal retrieval demonstrates that robust alignment often requires adaptive or self-supervised representation calibration~\cite{li2021self,huang2020explore}, suggesting potential degradation when query–evidence alignment is imperfect. 
Second, multi-turn retrieval and reasoning can be viewed as generating a trajectory of intermediate decisions; however, our method relies on local heuristics rather than explicitly optimizing a global trajectory-level objective. Recent advances in listwise and trajectory-level optimization indicate that directly aligning entire trajectories with final outcomes can better mitigate compounding errors~\cite{huang2025listwise}.
Exploring trajectory-aware objectives and noise-aware retrieval remains an important direction for future work.

\end{document}